\documentclass[epsfig,usenatbib]{mnras}
\usepackage{epsfig}
\usepackage{natbib}

\def\gtsima{$\; \buildrel > \over \sim \;$}
\def\ltsima{$\; \buildrel < \over \sim \;$}
\def\gtrsim{\lower.5ex\hbox{\gtsima}}
\def\lesssim{\lower.5ex\hbox{\ltsima}}

\begin{document}
%

%A nuclear star cluster surrounding ESO 243-49 HLX-1?
%The nature of ESO 243-49 HLX-1: the gas-rich minor merger scenario
%Is ESO 243-49 HLX-1 the remnant of a gas-rich minor merger?
%The origin of  ESO 243-49 HLX-1 from a gas-rich minor merger
%A minor merger scenario to explain ESO 243-49 HLX-1
\title[Massive black hole binaries from runaway collisions]{Massive black hole binaries from runaway collisions: the impact of metallicity}
\author[Michela Mapelli]
{Michela Mapelli$^{1,2}$
\\
$^1$INAF-Osservatorio Astronomico di Padova, Vicolo dell'Osservatorio 5, I--35122, Padova, Italy, {\tt michela.mapelli@oapd.inaf.it}\\
$^2$INFN, Milano Bicocca, Piazza della Scienza 3, I-20126 Milano, Italy\\
}
\maketitle \vspace {7cm }
\bibliographystyle{mnras}
 
\begin{abstract}
The runaway collision scenario is one of the most promising mechanisms to explain the formation of intermediate-mass black holes (IMBHs) in young dense star clusters. On the other hand, the massive stars that participate in the runaway collisions lose mass by stellar winds. In this paper, we discuss new N-body simulations of massive ($6.5\times{}10^4$ M$_\odot$) star clusters, in which we added upgraded recipes for stellar winds and supernova explosion at different metallicity. We follow the evolution of the principal collision product (PCP), through dynamics and stellar evolution, till it forms a stellar remnant. At solar metallicity, the mass of the final merger product spans from few solar masses up to $\sim{}30$ M$_\odot$. At low metallicity ($0.01-0.1$ Z$_\odot$) the maximum remnant mass is $\sim{}250$ M$_\odot$, in the range of IMBHs. A large fraction ($\sim{}0.6$) of  the PCPs are not ejected from the parent star cluster and acquire stellar or black hole (BH) companions. Most of the long-lived binaries hosting a PCP are BH-BH binaries. We discuss the importance of this result for gravitational wave detection.
\end{abstract}
\begin{keywords}
stars: kinematics and dynamics -- galaxies: star clusters: general -- stars: black holes -- gravitational waves -- methods: numerical -- stars: mass-loss
%stars: kinematics and dynamics -- galaxies: star clusters: general -- stars: black holes -- gravitational waves -- black hole physics -- stars: evolution -- methods: numerical -- stars: blue stragglers -- stars: mass-loss
\end{keywords}

%
%________________________________________________________________

\section{Introduction}
Young ($<100$ Myr) dense ($\gtrsim{}10^3$ M$_\odot{}$ pc$^{-3}$) star clusters (YDSCs) are one of the most crowded nurseries of stars in the local Universe \citep{lada2003}. Given its high central density, the core of a YDSC is the ideal environment for extreme dynamical interactions between stars, including stellar collisions, which are unlikely to occur in the galactic field (see \citealt{portegieszwart2010} for a review). Stellar collisions are enhanced when the core of the YDSC undergoes a gravothermal instability \citep{freitag2006b}, because the central density can grow by orders of magnitude. Most collisions are triggered by three-body encounters (i.e. close encounters between a star and a binary system, \citealt{heggie1975}), in the sense that they occur preferentially between one of the members of a binary system and a stellar intruder \citep{gaburov2008a}.
 
The most massive objects in a YDSC are more likely to undergo stellar collisions, because dynamical friction brings them to the YDSC core in few Myr \citep{gaburov2008a}.  Moreover, they are particularly efficient in forming binary systems through close encounters of three single stars and through dynamical exchanges \citep{hills1976,hills1980,hills1989,hills1991,hills1992,mcleod2015}. Thus, the most massive objects may experience multiple collisions \citep{portegieszwart1999}: the larger their mass becomes after a collision, the higher the probability that they participate in a further collision, unless they are dynamically ejected by the YDSC. According to the runaway collision scenario \citep{colgate1967,sanders1970,quinlan1990,ebisuzaki2001,portegieszwart1999,portegieszwart2002}, a massive object experiencing multiple collisions can become a very massive blue straggler star (with mass $>>100$ M$_\odot$) and then it might collapse to an  intermediate-mass black hole (IMBH), i.e. a black hole (BH) with mass larger than expected for stellar-mass BHs ($\sim{}10^2$ M$_\odot$, \citealt{spera2015}), but smaller than $\lesssim{}10^5$ M$_\odot$. % The runaway merger scenario is one of the principal mechanisms that were proposed to form IMBHs. Other popular scenarios of IMBH formation are the direct collapse of very massive ($>250$ M$_\odot$) population III stars \citep{heger2003,schneider2005}, and the repeated mergers of stellar-mass BHs with stars and other BHs, triggered by dynamical interactions in star clusters \citep{miller2002,giersz2015}.} 

The evolution of the collision product is one of the main open questions of  the runaway merger scenario. A very massive merger product may lose mass by stellar winds \citep{glebbeek2009}. At the end, the merger product may explode as a  supernova (SN, \citealt{portegieszwart2007,pan2012,heuvel2013}), or it may preserve enough mass to collapse to a BH directly \citep{fryer1999,heger2003,fryer2012,spera2015}. The mass of the final remnant is in the IMBH mass range only if mass loss by stellar winds is moderate and only if direct collapse takes place.

Several direct N-body simulations \citep{portegieszwart2002,moeckel2011} and Monte Carlo simulations \citep{gurkan2006,goswami2012} investigating the runaway merger do not include prescriptions for stellar evolution, mass loss by stellar winds and SN explosions. A number of studies adopt models for stellar evolution and/or approximate recipes for mass loss and SNe \citep{portegieszwart1999,portegieszwart2004,gurkan2004,freitag2006a,freitag2006b,arca2016}. Only few authors include a detailed treatment of mass loss during the collision \citep{gaburov2008b,gaburov2010} and afterward, as a consequence of stellar winds \citep{glebbeek2009,glebbeek2013,banerjee2012}. 

Hydro-dynamical simulations of the merger of massive stars show that up to $\sim{}25$ \% of the total mass of the colliding stars might be lost during the merger \citep{gaburov2010}. Moreover, all previous studies including recipes for stellar evolution agree that the massive merger product is expected to lose most of its mass by stellar winds at solar metallicity. Lower-metallicity environments may be the most important ones for the formation of IMBHs \citep{mapelli2009,belczynski2010,spera2015}, since stellar winds are less effective in metal-poor stars \citep{kudritzki1987,vink2001,kudritzki2002}. 

However, only few studies consider metal-poor clusters (e.g. \citealt{glebbeek2009}). Moreover, the evolutionary models of massive and very massive stars have been deeply revised in the last few years (e.g. \citealt{vink2001,vink2005,yungelson2008,tang2014,chen2015} and references therein). \cite{glebbeek2009} is one of the few studies that use up-to-date metallicity-dependent stellar evolution recipes (stellar winds are based on \citealt{vink2001} and \citealt{vink2005}). However, \cite{glebbeek2009} include these  stellar-wind recipes {\it a posteriori}, in  N-body simulations that were run without stellar evolution (or with older models of stellar evolution at solar metallicity). This procedure introduces a bias, since the mass loss of the merger product affects the probability that the merger product undergoes further collisions with other stars. Moreover, stellar winds affect the gravothermal instability phase of the host star cluster: they make the potential well of the core shallower, reducing the amount of kinetic energy that must be provided by three-body encounters to reverse the core collapse \citep{trani2014}. As a consequence, three-body encounters and stellar collisions are less effective, if stellar winds are included \citep{mapellibressan2013}. 

%If the runaway merger product 

%dynamics
%Most studies on the runaway merger scenario do not include stellar winds and SN recipes \citep{}. 

%(lack of) stellar evolution

In this paper, we study the runaway collision scenario through direct N-body simulations of YDSCs with different metallicity, adopting upgraded recipes for metallicity-dependent stellar evolution and mass loss by stellar winds \citep{mapelli2013}. %We focus on YDSCs whose initial conditions are less extreme than previously considered in the literature, their initial two-body relaxation time being ...... 
We investigate the evolution of the principal collision product (PCP), defined as the product of the first collision that occurs in a simulated YDSC, till it becomes a stellar remnant. Our main goal is to study whether (or not) the PCP acquires companions through dynamical processes and what are the properties of such PCP binaries. We focus on the importance of PCP binaries for the emission of gravitational waves (GWs) in YDSCs, in light of the recent first direct detection of GWs by Advanced LIGO \citep{abbott2016,LIGO2016,LIGO2016b}. %We show that massive BH binaries born from runaway collisions can play a key role among GW sources. This is an essential clue to interpret the recent first direct detection of GWs by Advanced LIGO \citep{detection and astrophysics papers}, and to make predictions for forthcoming detections.  

%aims of paper

 This paper is organised as follows. In Section~2, we describe the methods used in this paper and the simulation set up. Our results are presented in Section~3. In Section~4 we discuss the implications of our findings, focussing on GW detections. Our conclusions are summarized in Section~5. 
%gravitational waves

 %%%%%%%%%%%%%%%%%%%%%%%%%%%%%%%%%% TABLE 1%%%%%%%%%%%%%%%%%%%%%%%%%%%%%%%%%%%%%
\begin{table}
\begin{center}
\caption{\label{tab:table1}
Initial conditions of the $N-$body simulations.}
 \leavevmode
\begin{tabular}[!h]{ccccc}
\hline
$N_{\rm runs}$ &  $Z/$Z$_\odot$   & $N_{\ast}$ & $r_{\rm v}$/pc & $W_0$\\
\hline
10          & 1.0   & $10^5$    & 1            & 9\\
10          & 0.1  & $10^5$    & 1            & 9\\
10          & 0.01 & $10^5$    & 1            & 9\\ 
\noalign{\vspace{0.1cm}}
\hline
\end{tabular}
\begin{flushleft}
\footnotesize{Column 1: number of $N-$body realizations; column 2: metallicity; column 3: number of particles ($N_\ast$); column 4: virial radius $r_{\rm v}$; column 5: dimensionless central potential.}
\end{flushleft}
\end{center}
\end{table}
%%%%%%%%%%%%%%%%%%%%%%%%%%%%%%%%%%%%%%%%%%%%%%%%%%%%%%%%%%%%%%%%%%%%%%%%%%%%%%%%

\section{Methods: N-body simulations}\label{sec:methods}
\subsection{N-body and stellar evolution} 
We ran a set of 30 direct N-body simulations of YDSCs using the {\sc starlab} %\footnote{\tt http://www.sns.ias.edu/$\sim{}$starlab/} 
public software environment \citep{portegieszwart2001}. %, see also \citealt{portegieszwart1996,nelemans2001, anders2009}). 
{\sc kira}, the direct N-body integrator included in {\sc starlab}, implements a Hermite 4th order integration algorithm \citep{makino1992} and a neighbors--perturbers scheme to ensure an accurate integration of tight binaries and multiple  systems. We included stellar and binary evolution, using the modified version of {\sc starlab} described in \cite{mapelli2013}. Stars evolve in radius, temperature and luminosity at different metallicities,  based on the polynomial fitting formulae by \cite{hurley2000}. The original fitting formulae of \cite{hurley2000} were developed for stars with a maximum mass of 100 M$_\odot$. In our code, we extrapolate them for higher masses \citep{mapelli2013}. To avoid unphysical results, we request that the values of the radius of single stars are consistent with PARSEC stellar evolution tracks \citep{bressan2012,tang2014,chen2015} for stars with mass $>100$ M$_\odot$, as implemented in our new population synthesis code {\sc SEVN} \citep{spera2015}. 

A treatment of stellar winds is included both for main sequence (MS) and post-MS stars. Massive MS stars lose mass based on the metallicity-dependent fitting formulae described in \cite{vink2001}. The mass-loss rate of luminous blue variable (LBV) stars (i.e. stars whose  luminosity $L$ and radius $R$ satisfy the requirement that $L/{\rm L}_\odot{}>6 \times 10^5$ and $10^{-5}\,{}(R/{\rm R}_\odot{})\,{}(L/{\rm L}_\odot{})^{0.5} >1.0$, \citealt{HumphreysDavidson1994}) is $\dot{M} = f_{\rm LBV} \times 10^{-4}$ M$_\odot{}$ yr$^{-1}$, where $f_{\rm LBV}=1.5$ \citep{belczynski2010}. Wolf-Rayet (WR) stars undergo a mass-loss rate by stellar winds defined by $\dot{M} = 10^{-13} (L/{\rm L}_\odot{})^{1.5}\,{}({Z/{\rm Z}_\odot})^{\beta{}}$ M$_\odot{}$ yr$^{-1}$, where $\beta{}=0.86$ \citep{hamann1998,vink2005,belczynski2010}. Stellar winds in asymptotic giant branch (AGB) stars are modelled as in the standard version of {\sc starlab} \citep{portegieszwart1996}, and do not depend on metallicity.  

We assume that the mass lost by stellar winds and SNe is immediately removed from the simulation. This assumption is correct for SN ejecta and also for the winds of massive stars, which are expected to move fast ($\gtrsim{}1000-2000$ km s$^{-1}$,\citealt{vink2005,muijres2012}) with respect to the escape velocity of the simulated YDSCs ($\sim{}20-25$ km s$^{-1}$). %Stellar winds by AGB stars have much smaller velocities ($\approx{}10-20$ km s$^{-1}$, Loup et al. 1993), but still sufficiently high to escape from our simulated YDSCs. Furthermore, we show in Section~\ref{sec:result} that AGB stars do not play an important role for the results presented in this paper.

The formation of stellar remnants is implemented as described in \cite{mapelli2013}. In particular, BH masses at various metallicities follow the distribution described in fig. 1 of \cite{mapelli2013} (see also \citealt{fryer1999,fryer2001, fryer2012}). If the final mass  $m_{\rm fin}$  of the progenitor star (i.e. the mass bound to the star immediately before the collapse) is $>40$ M$_\odot$, we assume that the SN fails and that the star collapses quietly to a BH. The requirement that $m_{\rm fin}>40$ M$_\odot$ implies that only stars  with zero-age MS (ZAMS) mass $\gtrsim{}80$ and $\gtrsim{}100$ M$_\odot{}$, can undergo a direct collapse at $Z=0.01$ and 0.1 Z$_\odot{}$, respectively. 
%If $m_{\rm fin}\ge{}40$ M$_\odot{}$, the mass of the BH is derived as $m_{\rm BH}=m_{\rm CO}+f_{\rm coll}\,{}(m_{\rm He}+m_{\rm H})$, where $m_{\rm CO}$ is the final mass of the Carbon Oxygen (CO) content of the progenitor, while $m_{\rm He}$ and $m_{\rm H}$ are the residual mass of Helium (He) and of Hydrogen (H), respectively. $f_{\rm coll}$ is the fraction of He and H mass that collapses to the BH in the failed SN scenario.  The value of $f_{\rm coll}$ is uncertain. We assume $f_{\rm coll}=2/3$ to match the maximum values of $m_{\rm BH}$ at low $Z$ derived by \citep{belczynski2010}.
 The mass of a BH born from direct collapse is similar to the final mass of the progenitor star. Thus, BHs with mass  up to $\sim{}80$ M$_\odot$ ($\sim{}40$ M$_\odot$) can form if the metallicity of the progenitor is $Z\sim{}0.01$ Z$_\odot$ ($Z\sim{}0.1$ Z$_\odot$). \cite{spera2015} suggest that the threshold for direct collapse is even lower ($m_{\rm fin}\sim{}30$ M$_\odot$) and that the maximum mass of a BH for a given metallicity can be even higher than discussed in \cite{mapelli2013}, but these predictions cannot be implemented in our simulations because they still lack a treatment of binary evolution and its effects on the BH mass. %This simulation approach is suited for understanding the interplay between dynamical effects (three- and few-body encounters) and metallicity-dependent stellar evolution in YDSCs.

Neutron stars (NSs) are assumed to receive a natal kick drawn from the distribution of \cite{hartman1997}. BHs that form from quiet collapse are assumed to receive no natal kick \citep{fryer2012}. For BHs that form from a SN explosion, the natal kicks were drawn from the same distribution as NSs but scaled as $m_{\rm NS}/m_{\rm BH}$ (where $m_{\rm NS}=1.3$ M$_\odot{}$ is the typical mass of a NS, while $m_{\rm BH}$ is the mass of the considered BH), to preserve linear momentum. 

The evolution of binaries is implemented as described in \cite{portegieszwart1996} and \cite{portegieszwart2001}: we include mass transfer by stellar winds, mass transfer by Roche lobe overflow, magnetic braking, energy and angular-momentum loss due to GW emission \citep{peters1964}, tidal circularization, common envelope (following the formalism proposed by \citealt{webbink1984} and \citealt{dekool1990}, with binding energy coefficient $\lambda{}=0.5$ and efficiency coefficient $\alpha{}_{\rm CE}=1.0$).

If two stars approach each other by less than $d=(r_1+r_2)$, where $r_1$ and $r_2$ are the radius of the first and second star, respectively, they are merged to a single particle with mass equal to the total mass of the progenitor. This is an optimistic assumption, since it does not consider the relative velocity between the two stars and it assumes that no mass is lost during the merger. While on the MS, the merger product is rejuvenated by a rejuvenation fraction \citep{meurs1989,portegieszwart1999}
\begin{equation}\label{eq:frej}
f_{\rm rej}(m_1,m_2)=\frac{m_1}{m_1+m_2}\,{}\frac{\tau_{\rm MS}(m_1+m_2)}{\tau{}_{\rm MS}(m_1)},
\end{equation}
where $m_1$ and $\tau{}_{\rm MS}(m_1)$ are the mass and the MS lifetime of the most massive among the two colliding stars, $m_2$ is the mass of the other colliding star, and $\tau_{\rm MS}(m_1+m_2)$ is the MS lifetime of a star with mass equal to the mass of the collision product. The new age of the collision product is $t(m_1+m_2)=f_{\rm rej}(m_1+m_2)\,{}t(m_1)$, where $t(m_1)$ is the age of the first progenitor star at the time of the collision.  For collisions of post-MS stars the rejuvenation is calculated with a similar formula, in which we substitute $\tau{}_{\rm MS}$ with the lifetime of the corresponding evolutionary stage (e.g. $\tau{}_{\rm SG}$ for a sub-giant star), while $t(m_1)$ indicates the time the star of mass $m_1$ has already spent on its current evolutionary stage (e.g. the time spent in the sub-giant branch), instead of the total stellar age (see \citealt{portegieszwart2001} for details). Rejuvenation of a MS star is generally  more conspicuous than rejuvenation of post-MS stars, since the time a star spends in post-MS evolutionary stages is $<1/10$ of its entire lifetime.  %and because of the uncertainties in the treatment of post-collision mixing (e.g. \citealt{glebbeek2013}).}
%no rejuvenation is assumed. We can neglect this ingredient for the purposes of this work, since the time a star spends in post-MS stages is $<1/10$ of its entire lifetime and because of the uncertainties in the treatment of post-collision mixing (e.g. \citealt{glebbeek2013}).}

%{\bf For collisions of post-MS stars no rejuvenation is assumed. We can neglect this ingredient for the purposes of this work, since the time a star spends in post-MS stages is $<1/10$ of its entire lifetime and because of the uncertainties in the treatment of post-collision mixing (e.g. \citealt{glebbeek2013}).}
%Adding corrections for the rejuvenation of post-MS stars requires a dedicated study, given the uncertainties on post-collision mixing \citep{}. However, we can neglect it for the purposes of this work, since the time a star spends in post-MS stages is $<1/10$ of its entire lifetime.} %(the lifetime on post-MS evolutionary stages is very short if compared to the integration timesteps).

%%%%%%%%%%%%%%%%%%%%%%%%%%%%%%%%%%% FIGURE 1 %%%%%%%%%%%%%%%%%%%%%%%%%%%%%%%%%%
\begin{figure}
\center{{
\epsfig{figure=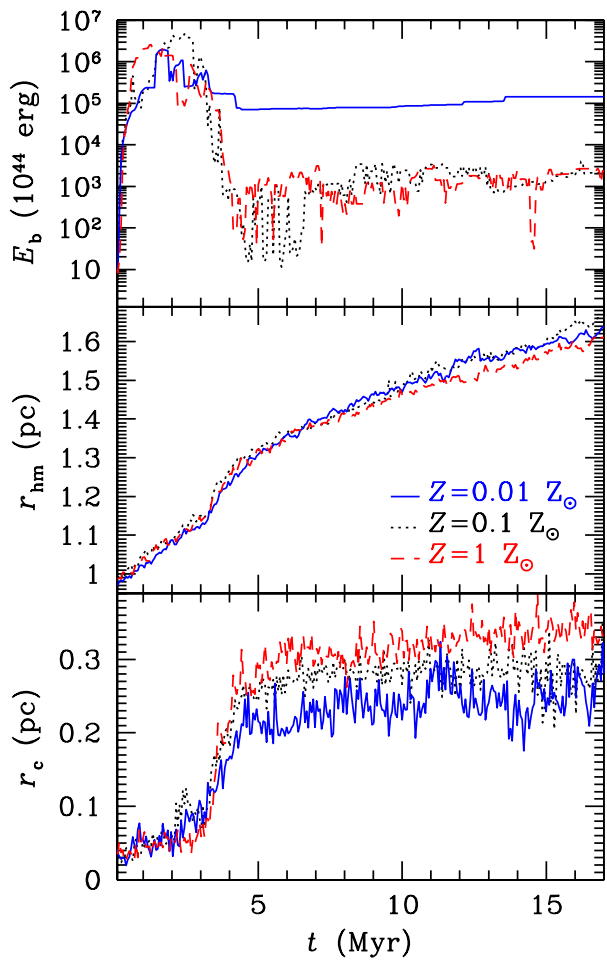,width=8.5cm} %was fig0.eps
}}
\caption{\label{fig:fig1}
 Bottom panel: evolution of the core radius $r_{\rm c}$ as a function of time; central panel: evolution of the half-mass radius $r_{\rm hm}$ as a function of time; top panel: evolution of the total binding energy stored in binaries $E_{\rm b}$ as a function of time. Each line is the median value of 10 realizations of the same YDSC model. Red dashed line: $Z=1$ Z$_\odot$; black dotted line: $Z=0.1$ Z$_\odot$; blue solid line: $Z=0.01$ Z$_\odot$. 
}
\end{figure}
%%%%%%%%%%%%%%%%%%%%%%%%%%%%%%%%%%%%%%%%%%%%%%%%%%%%%%%%%%%%%%%%%%%%%%%%%%%%%%%

\subsection{Initial conditions and simulations grid}
We simulate 30 YDSCs with $N=10^5$ particles each (in our simulations each particle is a single star), corresponding to an average total mass $M_{\rm TOT}\sim{}6.5\times{}10^4$ M$_\odot{}$. The YDSCs are modelled as King models \citep{king1966} with virial radius $r_{\rm v}=1$ pc, core radius $r_{\rm c}\sim{}0.05$ pc, half-mass radius $r_{\rm hm}\sim{}0.98$ pc, and central dimensionless potential $W_0=9$ (Fig.~\ref{fig:fig1}). These values are reminiscent of the YDSCs in the Milky Way \citep{portegieszwart2010}, and are not as extreme as the ones adopted in previous studies (e.g. \citealt{gaburov2008a}). We generate star masses according to a Kroupa initial mass function (IMF, \citealt{kroupa2001}) with minimum mass $m_{\rm low}=0.1$ M$_\odot$ and maximum mass $m_{\rm up}=150$ M$_\odot$. %\footnote{\bf The original fitting formulae of \cite{hurley2000} were developed for stars with a maximum mass of 100 M$_\odot$. In our code, we extrapolate them for higher masses \citep{mapelli2013}. To avoid unphysical results, we request that the values of the radius of single stars are always consistent with PARSEC stellar evolution tracks \citep{bressan2012,cheng2014,tang2015} for stars with mass $>150$ M$_\odot$, as implemented in our new population synthesis code SEVN \citep{spera2015}.} $m_{\rm up}=150$ M$_\odot$. 

We do not include primordial binaries. These are the most computationally expensive ingredient, but should be accounted for in future work, because a high binary fraction is observed in young star clusters (e.g. \citealt{li2013}). If there are no primordial binaries, binaries can form through  close encounters of three single stars, during which two stars transfer sufficient kinetic energy to the third one to become bound to each other, while the third star flies away.

%Actually, since we adopt a strong concentration (dimensionless central potential $W_0=9$) and since  the initial conditions of the simulated star clusters are close to core collapse (Fig.~\ref{fig:fig1}), and binaries immediately form by dynamical encounters\footnote{\bf This process has already been discussed by \cite{trani2014}. Here, we just summarize the main points, and we refer to \cite{trani2014} for more details. Given two identical N-body models of star clusters except for the central dimensionless potential $W_0$, the model with the higher $W_0$ will start core collapse before the other one. Figures 3 and 5 of \cite{trani2014} show this trend: models with $W_0=5$ (Figure 3) undergo core collapse at $3$ Myr (i.e. $\approx{}0.2\,{}t_{\rm hm}$, where $t_{\rm hm}$ is the half-mass relaxation time), while models with $W_0=9$ (Figure 5) enter core collapse in the first Myr. Binaries start to form efficiently during core collapse, thus at $t\sim{}3$ Myr and at $t\sim{}0$ Myr if $W_0=5$ and $W_0=9$, respectively (see the right-hand panels of fig.~3 and 5 of \citealt{trani2014}).}.

We consider star clusters at three different metallicities: $Z=0.02,$ 0.002 and 0.0002. Assuming $Z_\odot{}=0.02$, this means that we consider star clusters with metallicity $Z=$ 1, 0.1, and 0.01 Z$_\odot$, respectively. We simulate 10 different realizations for each of these metallicities (Table~\ref{tab:table1}), to  filter out stochastic fluctuations.  

 We  simulate each YDSC for $t=17$ Myr. At this time, the stage of runaway merger has already completed and binaries including BHs have already formed.

In our simulations, we do not include the contribution of the tidal field of the host galaxy, because this would add more parameters to deal with. Milky Way YDSCs are observed both near the Galactic centre (Arches, Quintuplet) and close to the Sun (Trumpler~14, \citealt{portegieszwart2010}). Accounting for various tidal fields would require a much larger sample of simulations. %In the top panel of Fig.~\ref{fig:fig1}, we can see that most binaries form in the first $\sim{}5$ Myr. After this stage, existing hard binaries (i.e. binaries with binding energy larger than the average kinetic energy of a star in the star cluster, \citealt{heggie1975}) do not change dramatically (see also \citealt{ziosi2014}).
 %In absence of tidal fields, the simulated YSC could evolve forever without significant changes (except for mass loss by asymptotic giant branch stars and the increase of the half-mass radius). 
Moreover, the Galactic tidal field is not sufficiently strong to disrupt the YDSCs for the entire duration of our simulations (17 Myr), even for YDSCs close to the Galactic centre \citep{gieles2011}. %In a forthcoming study, we will include the effect of tidal fields.} 

\section{Results}\label{sec:results} 

\subsection{Structural evolution of the YDSCs and formation of binaries}\label{sec:structure}
 Fig.~\ref{fig:fig1} shows the evolution of the core radius $r_{\rm c}$ (bottom panel) and the half-mass radius $r_{\rm hm}$ (central panel) of the simulated YDSCs as a function of time for the three different metallicities (each line is the median value of $r_{\rm c}$ and $r_{\rm hm}$ over 10 runs with the same metallicity, to filter our stochastic fluctuations). Both the core and the half-mass radius start expanding rapidly at time $t\sim{}3$ Myr.

The impact of mass loss by stellar winds and SNe on the evolution of $r_{\rm c}$ and $r_{\rm hm}$ has already been discussed in \cite{mapellibressan2013} and \cite{trani2014}. Here, we recall that massive stars ($<30$ M$_\odot$) lose most of their mass by stellar winds in the first $\sim{}2-6$ Myr, while most SNe occur between $\sim{}3$ and $\sim{}50$ Myr. At low-metallicity ($Z\le{}0.1$ Z$_\odot$) the most massive stars are allowed to collapse to BH without SN. 

The mass ejected by stellar winds and SNe is lost from the YDSC, making the potential well shallower. Thus, mass loss contributes to drive the expansion of the core (Fig.~\ref{fig:fig1}). Three-body encounters, which transfer kinetic energy to stars, are the other main driver of the expansion of the core. The contribution of stellar winds is more important at high metallicity; in fact, the core radius of metal-rich clusters expands more than that of metal-poor clusters (Fig.~\ref{fig:fig1}).   
%as shown by the fact that the core radius of metal-rich clusters expands more than that of metal-poor clusters (Fig.~\ref{fig:fig1}). 

The top panel of Fig.~\ref{fig:fig1} shows the total binding energy of binaries $E_{\rm b}$, i.e. the sum of the binding energies of all binaries in a YDSC at a given time. 
%, normalized to the average kinetic energy of a star in the YDSC ($k_{\rm B}$ T, where $k_{\rm B}$ is the Boltzmann constant and T is the `kinetic' temperature), as a function of time. %In particular, each line is the median value of $E_{\rm b}$ over 10 simulations with the same metallicity. 
The binding energy of a binary in Fig.~\ref{fig:fig1} drops to zero only if the binary merges or becomes unbound (because of SN kicks or three-body encounters). Binaries that are ejected from the YDSC are not removed from Fig.~\ref{fig:fig1}.  

Binaries form very rapidly at the beginning of the simulations, regardless of the metallicity. The formation of binaries is driven  by encounters of three single stars.

The reason why binary formation is so efficient in the first $\sim{}5$ Myr can be explained as follows. The initial half-mass relaxation timescale of our YDSCs is \citep{portegieszwart2010} 
\begin{equation}
t_{\rm rlx}\sim{}50\,{}{\rm Myr}\left(\frac{M_{\rm TOT}}{6.5\times{}10^4\,{}{\rm M}_\odot}\right)^{1/2}\,{}\left(\frac{r_{\rm v}}{1\,{}{\rm pc}}\right)^{3/2}\,{}\left(\frac{\langle{}m\rangle{}}{1\,{}{\rm M}_\odot}\right)^{-1},
\end{equation} 
where $M_{\rm TOT}$ is the total mass of the YDSC, $r_{\rm v}$ is the virial radius, and $\langle{}m\rangle{}$ is the average mass of a star. Thus, the dynamical friction timescale of a star of mass $m$ is
\begin{equation}
t_{\rm DF}(m)\sim{}\frac{\langle{}m\rangle{}}{m}\,{}t_{\rm rlx}\sim{}2\,{}{\rm Myr}\left(\frac{m}{25\,{}{\rm M}_\odot}\right)^{-1}.
%t_{\rm DF}(m)\sim{}t_{\rm rlx}\,{}\,{}\left(\frac{\langle{}m\rangle{}}{m}\right)\sim{}2\,{}{\rm Myr}\left(\frac{m}{25\,{}{\rm M}_\odot}\right)^{-1}.
\end{equation} 
This means that the massive stars segregate to the core of the YDSC in few Myr. Thanks to the high density of the core, the massive stars interact with each other and effectively build binaries, before undergoing SN explosion or directly collapsing to a BH.

At $t>5$ Myr, the binary binding energy of YDSCs with metallicity $Z=0.01$ Z$_\odot$ is much higher (more than one order of magnitude) than that of YDSCs with $Z=0.1,$ 1 Z$_\odot$. %The difference between the binding energy evolution at $Z=0.01$ Z$_\odot$ with respect to the higher metallicity becomes very important (more than one order of magnitude) at $t>5$ Myr. 
%Even if a single very energetic binary is sufficient to affect the median value in Fig.~\ref{fig:fig1}, 
This large difference comes from the fact that metal-poor binaries can be significantly more massive at late epochs, because they lose less mass by stellar winds and collapse into more massive BHs than metal-rich systems. Moreover, binaries tend to live for a longer time at low metallicity, because BHs born from direct collapse receive no natal kick, and, if they are members of a binary, the binary is not destroyed by the kick.  %that most metal-poor massive stars collapse to BH directly, avoiding SN explosion. In this case, the BH forms without natal kick and, if the BH  is member of a binary,  the binary is not destroyed. }

%This large difference comes from the fact that the binaries can be significantly more massive at $Z=0.01$ Z$_\odot$ than at higher metallicity, after massive stars have evolved into BHs. Moreover, binaries tend to live for a longer time at low metallicity. The main reason is that most metal-poor massive stars collapse to BH directly, avoiding SN explosion. In this case, the BH forms without natal kick and, if the BH  is member of a binary,  the binary is not destroyed.
%At $\sim{}3-4$ Myr, the core collapse is over in all runs, and the half-mass radius keeps expanding. After this stage, the central density decreases and stellar collisions become unlikely. 

\subsection{Mass evolution of principal collision products (PCPs)}
 In each simulated YDSC, we identify the first two stars that have a genuine collision (we exclude mergers in contact binaries triggered by stellar evolution) and we track the subsequent history of this first collision product (hereafter, the principal collision product, PCP). Thus, we have a sample of 10 PCPs for each simulated metallicity (in Appendix~A, we also examine some of the properties of the other collision products). 

According to our definition, we consider as `genuine collisions' all collisions that are not triggered uniquely by stellar or binary evolution. Thus, the collision might occur between two single stars, or between a single star and the member of a binary, or between two members of a binary that was dynamically perturbed by the dynamical encounter with a third body. We find no PCP born from the collision of two single stars. All PCPs originate from encounters that involve a binary, confirming that binary interactions are very important for the runaway merger \citep{gaburov2008a}. Most of PCPs form from the collision between a member of a binary and a third star in a three-body encounter (70, 60, and 50\% at $Z=1,$ 0.1 and 0.01 Z$_\odot{}$, respectively). The remaining PCPs form from the collision between the two former members of a binary, and the collision is triggered by a three-body encounter. 

%%%%%%%%%%%%%%%%%%%%%%%%%%%%%%%%%%% FIGURE 2 %%%%%%%%%%%%%%%%%%%%%%%%%%%%%%%%%%
\begin{figure}
\center{{
\epsfig{figure=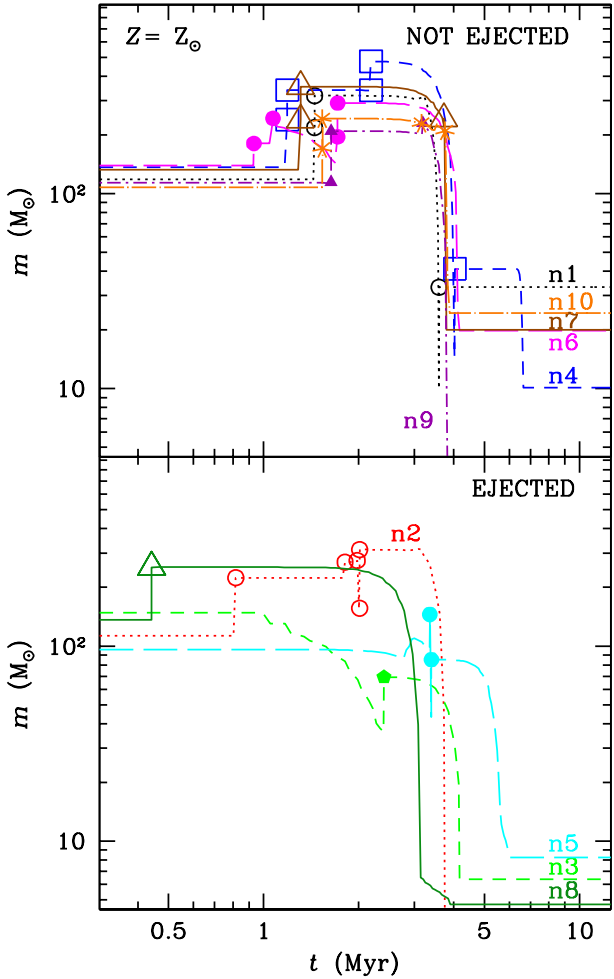,width=8.5cm} %newPCPmass_Z1.eps
}}
\caption{\label{fig:fig2}
Mass evolution of PCPs in runs with metallicity $Z=$ Z$_\odot$.  Top panel: PCPs that are not ejected from the YDSC.  Bottom panel: PCPs that are ejected from the YDSC. Each line corresponds to the evolution of a single PCP  till the formation of the stellar remnant. Symbols mark single collisions in the evolution of a PCP. Dotted black line (with open circles): run n1; dotted red line (with open circles): run n2; dashed green line (with filled pentagons): run n3; dashed blue line (with open squares): run n4; long-dashed cyan line (with filled circles): run n5; long-dashed magenta line (with filled circles): run n6; solid brown line (with open triangles): run n7; solid dark green line (with open triangles): run n8; dot-dashed violet line (with filled triangles): run n9; dot-long-dashed orange line (with asterisks): run n10.
}
\end{figure}
%%%%%%%%%%%%%%%%%%%%%%%%%%%%%%%%%%%%%%%%%%%%%%%%%%%%%%%%%%%%%%%%%%%%%%%%%%%%%%%
%%%%%%%%%%%%%%%%%%%%%%%%%%%%%%%%%%% FIGURE 3 %%%%%%%%%%%%%%%%%%%%%%%%%%%%%%%%%%
\begin{figure}
\center{{
\epsfig{figure=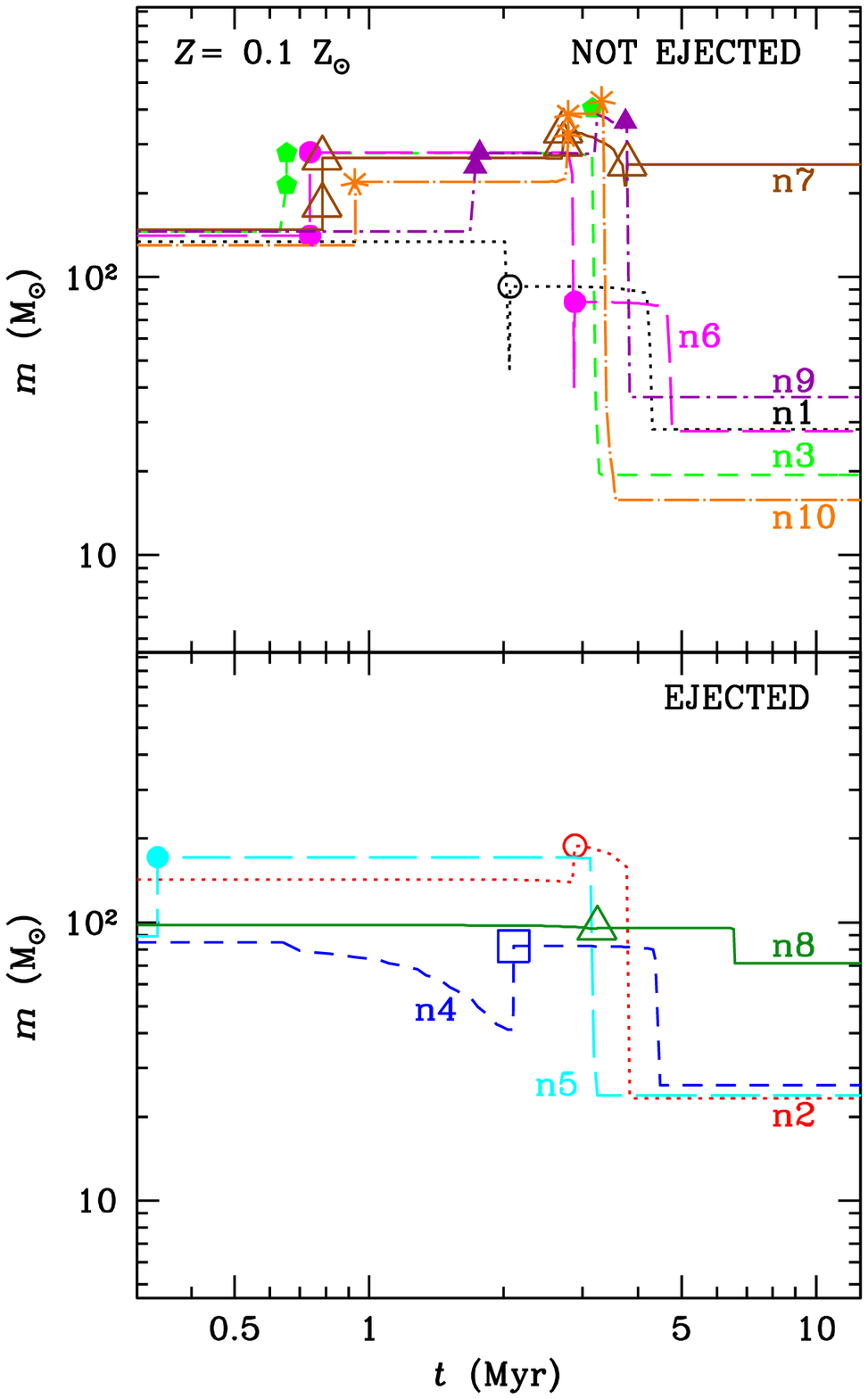,width=8.5cm} %was newPCPmass_Z01.eps
}}
\caption{\label{fig:fig3}
 Same as Fig.~\ref{fig:fig2} but for metallicity $Z=0.1$ Z$_\odot$. %Each line corresponds to the evolution of a single PCP  till the formation of the stellar remnant. Dotted black line (with open circles): run n1; dotted red line (with open circles): run n2; dashed green line (with open squares): run n3; long-dashed blue line (with filled circles): run n4; dashed cyan line (with open squares): run n5; long-dashed magenta line (with filled circles): run n6; solid brown line (with open triangles): run n7; solid dark green line (with open triangles): run n8; dot-dashed violet line (with filled triangles): run n9; dot-long-dashed orange line (with asterisks): run n10.
}
\end{figure}
%%%%%%%%%%%%%%%%%%%%%%%%%%%%%%%%%%%%%%%%%%%%%%%%%%%%%%%%%%%%%%%%%%%%%%%%%%%%%%%

%%%%%%%%%%%%%%%%%%%%%%%%%%%%%%%%%%% FIGURE 4 %%%%%%%%%%%%%%%%%%%%%%%%%%%%%%%%%%
\begin{figure}
\center{{
\epsfig{figure=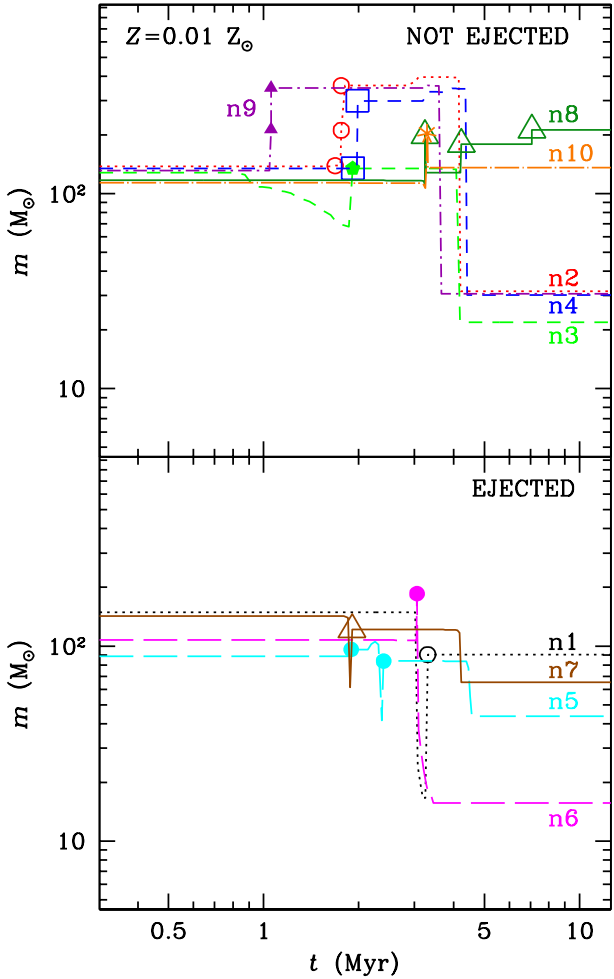,width=8.5cm} %was newPCPmass_Z001.eps
}}
\caption{\label{fig:fig4}
 Same as Fig.~\ref{fig:fig2} but for metallicity $Z=0.01$ Z$_\odot$. %Each line corresponds to the evolution of a single PCP  till the formation of the stellar remnant. Dotted black line (with open circles): run n1; dotted red line (with open circles): run n2; dashed green line (with open squares): run n3; long-dashed blue line (with filled circles): run n4; dashed cyan line (with open squares): run n5; long-dashed magenta line (with filled circles): run n6; solid brown line (with open triangles): run n7; solid dark green line (with open triangles): run n8; dot-dashed violet line (with filled triangles): run n9; dot-long-dashed orange line (with asterisks): run n10.
}
\end{figure}
%%%%%%%%%%%%%%%%%%%%%%%%%%%%%%%%%%%%%%%%%%%%%%%%%%%%%%%%%%%%%%%%%%%%%%%%%%%%%%%
Figures~\ref{fig:fig2}, \ref{fig:fig3} and \ref{fig:fig4} show the mass evolution of the PCP at the three considered metallicities, distinguishing between PCPs that are ejected before the end of the simulation and PCPs that are not. Symbols in these Figures mark the time of each collision. In all set of runs, $40$\% of the PCPs are ejected from the parent YDSC. %, as a consequence of dynamical interactions. 
The retained PCPs undergo on average $\sim{}2.9$ collisions during the entire simulations, whereas the ejected PCPs undergo only $\sim{}1.5$ collisions, because no collisions occur after they are ejected (we note however that in run n2 at  $Z=$ Z$_\odot$ the PCP undergoes 5 collisions before being ejected). 

The average number of collisions is lower than previously reported by several authors \citep{portegieszwart1999,glebbeek2009}. The main reasons are that (i) we include  the contribution of stellar winds, which quench the core collapse and keep the central stellar density lower \citep{trani2014,mapellibressan2013}, and (i) our simulated YDSCs have less extreme central densities than those assumed in previous studies. For example, the virial radii  considered in \cite{gaburov2008a} span from 0.05 to 0.75 times those assumed in our simulated YDSCs. %but are similar to the YDSCs observed in the Milky Way

At solar metallicity (Fig.~\ref{fig:fig2}) the final mass of the PCP is always low ($\lesssim{}30$ M$_\odot$), because the PCP undergoes strong mass loss by stellar winds. For example, the star that undergoes the largest number of collisions at $Z=$ Z$_\odot$ (run n4) starts from a ZAMS mass of $\sim{}136$ M$_\odot$, collides for the first time with a binary system at $t\sim{}1.2$ Myr, before mass loss by stellar winds becomes important. Its mass becomes $\sim{}340$ M$_\odot$ and the object is rejuvenated according to equation~\ref{eq:frej}. It undergoes a second collision at time $t\sim{}2.16$ Myr, with another binary system (collisions with binary systems are frequent, because binary systems have a larger cross-section than single stars), and its mass rises  up to $\sim{}476$ M$_\odot$. Right after this collision, stellar winds start to be important and the mass of the object drops fast, till it collides with one more object (another O-type object, with mass $\sim{}41$ M$_\odot$). The final merger product loses most of its mass by stellar winds, undergoes a SN explosion and becomes  a $\sim{}10$ M$_\odot$ BH.

At lower metallicity (Figs.~\ref{fig:fig3} and \ref{fig:fig4}) the situation is different, because stellar winds are less efficient. For example, the star that undergoes the largest number of collisions at  $Z=0.1$ Z$_\odot$ (run n7) starts from a ZAMS mass of $\sim{}148$ M$_\odot$. It collides for the first time at $t=0.8$ Myr (well before stellar winds become important) with a binary system, and reaches a mass of $\sim{}268$ M$_\odot$. It is rejuvenated according to equation~\ref{eq:frej}. At $\sim{}2.7$ Myr it undergoes a further collision with another binary, and the mass of the PCP becomes $\sim{}334$ M$_\odot$. At this stage, the PCP is a very massive O-type star with an age $\sim{}3$ Myr: it suffers from stellar winds even if it is relatively metal-poor (it is a radiation pressure dominated object, anyway). At the end of the mass-loss phase it becomes a BH of $\sim{}210$ M$_\odot{}$ by direct collapse.

At $t\sim{}3.8$ Myr, the BH collides one more time with a hyper-giant star (with mass $\sim{}40$ M$_\odot$). According to our simplified recipes, the entire mass of the hyper-giant star is swallowed by the BH, which becomes a $\sim{}254$ M$_\odot$ BH. A BH of $\sim{}254$ M$_\odot$ can be considered an IMBH. 

We stress that a metallicity $Z=0.1$ Z$_\odot$ is low, but not dramatically so: this is the typical metallicity of several dwarf galaxies in the local Universe \citep{mapelli2010}. We note that the most massive BHs in Figs.~\ref{fig:fig3} and ~\ref{fig:fig4} come from PCPs that undergo the last collision relatively late, when they would have started losing mass by stellar winds if they had not suffered a new collision.  However, not all PCPs that undergo the last collision relatively late succeed in forming a massive BH, because the final mass of the BH depends on several ingredients (stellar winds, SN explosion) other than the dynamical encounters.

 One important caveat about Figs.~\ref{fig:fig2}, \ref{fig:fig3} and \ref{fig:fig4} is that the mass evolution of the PCP is sensitive to many physical processes, including (1) the rejuvenation  induced by the merger, (2) mass transfer to a companion star,  (3) the evolutionary stage of the colliding objects (MS or post-MS), and (4) the amount of fallback to the final BH.

 There are several uncertainties and approximations in our treatment of these processes. For example, the rejuvenation of the merger product should be estimated  accounting for the actual chemical mixing \citep{gaburov2008b}. Moreover, it is unlikely that no mass is lost during the collision \citep{gaburov2010}. Thus, the results depicted in Figs.~\ref{fig:fig2}, \ref{fig:fig3} and \ref{fig:fig4} should be regarded as rather optimistic.  %Forthcoming studies are necessary to improve this approach, and to reach a complete understanding of the evolution of collision products in YDSCs. }

%%%%%%%%%%%%%%%%%%%%%%%%%%%%%%%%%%% FIGURE 5 %%%%%%%%%%%%%%%%%%%%%%%%%%%%%%%%%%
\begin{figure}
\center{{
\epsfig{figure=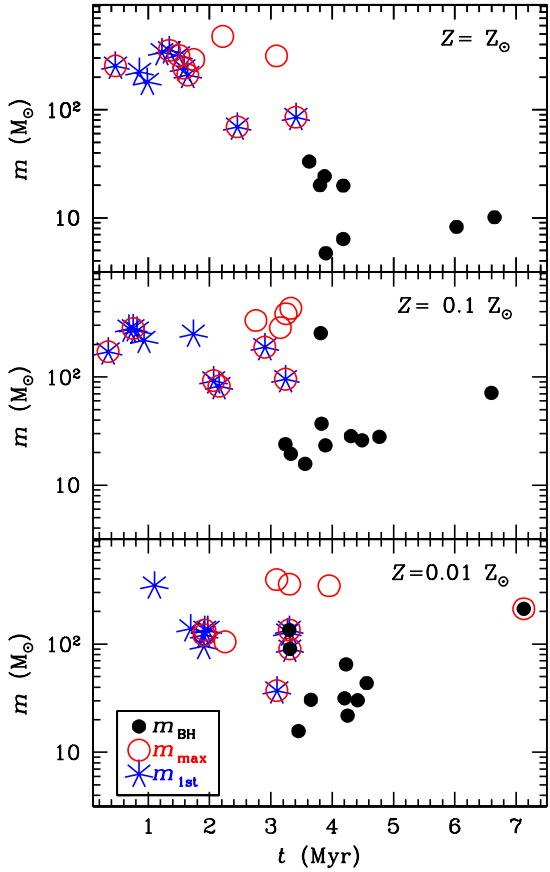,width=7.5cm} %was maxmassBH.eps 
}}
\caption{\label{fig:fig5}
Mass of the PCP after the first collision ($m_{\rm 1st}$, blue asterisks), maximum mass of the PCP ($m_{\rm max}$, red open circles) and mass of the BH generated by the PCP ($m_{\rm BH}$, black filled circles) as a function of time, at solar metallicity (top), $Z=0.1$ Z$_\odot$ (center) and $Z=0.01$ Z$_\odot$ (bottom). 
}
\end{figure}
%%%%%%%%%%%%%%%%%%%%%%%%%%%%%%%%%%%%%%%%%%%%%%%%%%%%%%%%%%%%%%%%%%%%%%%%%%%%%%%

Finally, Fig.~\ref{fig:fig5} is a summary of the most important properties of the PCPs: for each PCP at each metallicity we show the mass after the first collision, the maximum mass ever reached, and the mass of the final remnant. In several cases, the mass reached after the first collision is also the maximum mass ever, because the PCP undergoes only one collision and/or because stellar winds prevent larger masses to be reached later on. For most PCPs, the maximum mass is $>100$ M$_\odot$. In the most extreme cases $m_{\rm max}\sim{}300-500$ M$_\odot$. This result is relevant for the observation of very massive stars in the local Universe (e.g. in the R136 star cluster, \citealt{crowther2010}).

The mass of the remnant is the same as the maximum mass of the PCP only in three cases, all of them at $Z=0.01$ Z$_\odot{}$. At $Z=$ Z$_\odot$ all BH masses are in the $\sim{}5-30$ M$_\odot$ range (there are even two objects that do not become BHs). At $Z=0.1$ Z$_\odot$ most BH masses are between $\sim{}10$ and $\sim{}40$ M$_\odot$, which is consistent with the BH mass we expect at this metallicity, given the stellar evolution recipes we adopt \citep{mapelli2013}. However, there are two objects with mass 71 and 254 M$_\odot$, respectively, which can be considered in the IMBH mass range. At $Z=0.01$ Z$_\odot{}$, BH masses are between $\sim{}10$ and $\sim{}70$ M$_\odot$, which is, also in this case, the mass range that we expect at this metallicity, given the stellar evolution recipes we adopt. However, three BHs have mass above this range (90, 135 and 212 M$_\odot$), and can be considered IMBHs.

%\subsection{Statistics of PCPs}
%%%%%%%%%%%%%%%%%%%%%%%%%%%%%%%%%%% FIGURE 6 %%%%%%%%%%%%%%%%%%%%%%%%%%%%%%%%%%
\begin{figure}
\center{{
\epsfig{figure=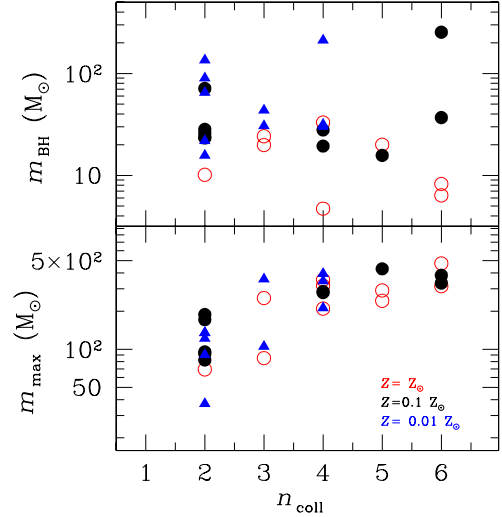,height=7.5cm} %was maxbh_ncoll.eps
}}
\caption{\label{fig:fig6}
Number of stars that collide ($n_{\rm coll}$) versus the maximum PCP mass (bottom panel) and versus the mass of the BH born from the PCP (top panel).  Red open circles: $Z=1$ Z$_\odot{}$; black filled circles: $Z=0.1$ Z$_\odot{}$; blue filled triangles: $Z=0.01$ Z$_\odot$. 
}
\end{figure}
\begin{figure}
\center{{
\epsfig{figure=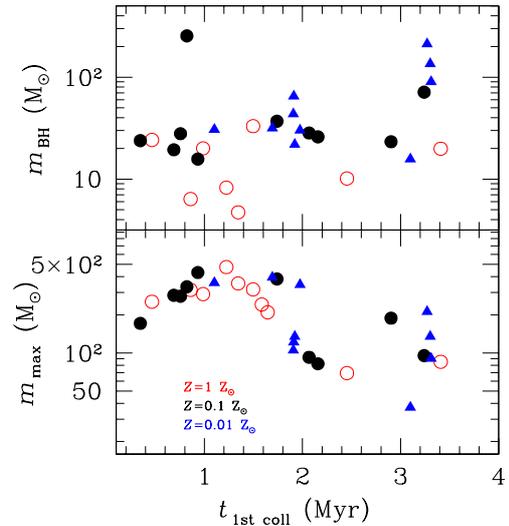,height=7.5cm} %was maxbh_1stcoll.eps
}}
\caption{\label{fig:fig7}
Time of the first collision ($t_{\rm 1st\,{}coll}$) versus the maximum PCP mass (bottom panel) and versus the mass of the BH born from the PCP (top panel). Red open circles: $Z=1$ Z$_\odot{}$; black filled circles: $Z=0.1$ Z$_\odot{}$; blue filled triangles: $Z=0.01$ Z$_\odot$. 
}
\end{figure}

%%%%%%%%%%%%%%%%%%%%%%%%%%%%%%%%%%%%%%%%%%%%%%%%%%%%%%%%%%%%%%%%%%%%%%%%%%%%%%%
Fig.~\ref{fig:fig6} shows the number of stars that collide to form a PCP versus the maximum PCP mass (bottom panel) and versus the mass of the BH born from the PCP (top panel). In the bottom panel, there is a clear trend: the more stars collide, the higher the maximum mass of the PCP can be. There is a similar trend between the maximum PCP mass and the time when the first collision occurred (Fig.~\ref{fig:fig7}, bottom panel): the sooner the PCP starts building, the more massive it can become.

However, this trend does not persist if we look at the BH mass: the mass of the BH that forms from the PCP depends on neither the number of stars that collide (Fig.~\ref{fig:fig6}, top panel) nor the time of the first collision (Fig.~\ref{fig:fig7}, top panel). This confirms that the mass of the remnant depends on the properties of the late evolution of the PCP (e.g. when the last collision occurs) rather than on the first collision and on the number of collisions. Overall, the masses of BHs born from PCPs span a large range of values, regardless of the number and time of collisions. This happens because the BH mass depends on many ingredients (stellar evolution, stellar winds and SN model) other than the dynamical evolution of the progenitor \citep{fryer2012,mapelli2013,spera2015}. 

%%%%%%%%%%%%%%%%%%%%%%%%%%%%%%%%%%% FIGURE 7 %%%%%%%%%%%%%%%%%%%%%%%%%%%%%%%%%%
%%%%%%%%%%%%%%%%%%%%%%%%%%%%%%%%%%%%%%%%%%%%%%%%%%%%%%%%%%%%%%%%%%%%%%%%%%%%%%%

%\subsection{Life in binary of the PCP}
\subsection{PCP binary systems}
In this Section, we study whether the PCP acquires a companion and evolves in a binary system. This aspect has not been highlighted in previous papers, but is extremely important for several reasons. First, the most massive stars in the local Universe are often found in binary systems \citep{crowther2010}. Moreover, if the PCP ends its life as a BH that is more massive than most of the other BHs in the YDSC, its efficiency in acquiring companions (and especially compact-object companions) is  higher: this is  important for GW observations.

\begin{figure}
\center{{
\epsfig{figure=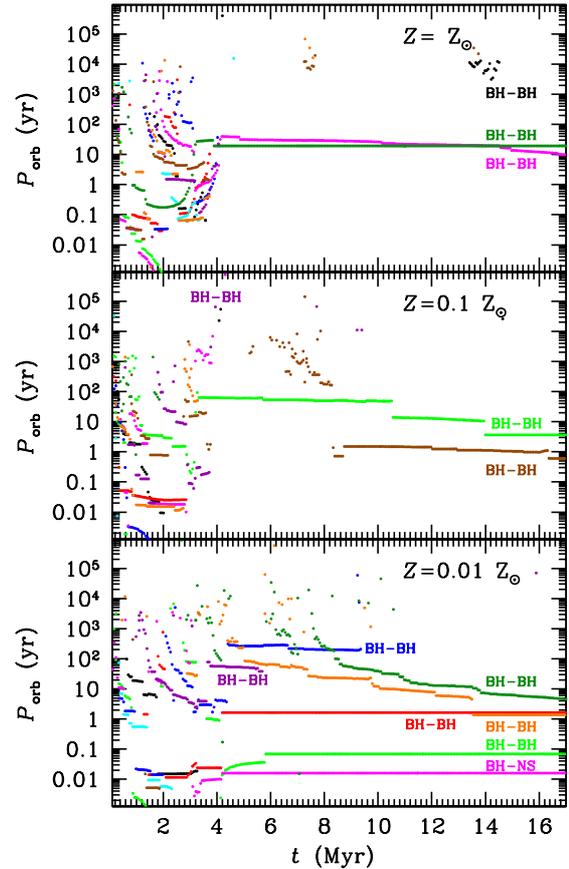,width=7.5cm} %was PCP_period.eps
}}
\caption{\label{fig:fig8}
Period of the binary systems whose member is a PCP as a function of time. From top to bottom: $Z=$ Z$_\odot$, $Z=0.1$ Z$_\odot$ and $Z=0.01$ Z$_\odot$. Each line is a single PCP. The labels 'BH-BH' and 'BH-NS' mark PCPs that are members of 'black hole-black hole' and 'black hole-neutron star' binaries, respectively. Colours associated with the single runs are the same as in Figs.~\ref{fig:fig2}, \ref{fig:fig3}, and \ref{fig:fig4}.
}
\end{figure}

Fig.~\ref{fig:fig8} shows the period of the binary hosting the PCP (when the PCP is member of a binary) as a function of time, for the entire duration of the simulations. During the early stages ($t<4$ Myr), the situation  is extremely complex. At the beginning of the simulations, the density of the YDSC is maximum (Fig.~\ref{fig:fig1}). The PCP, which is still building its mass through collisions, acquires companions dynamically: orbital periods $>>10$ yr correspond to binaries that formed dynamically, often with a large eccentricity, and then become more and more bound via three-body encounters (as we see from the fact that the period decreases if the binary system survives). Some binary systems reach an extremely short period ($<<0.1$ yr) and then disappear from the plot: these are mergers of the PCP with its companion stars.  Each PCP undergoes $\sim{}1$ merger, on average, with one of its companion stars during the entire simulation; most of the mergers occur in the first 4 Myr, when both the PCP and the companion have not collapsed to dark remnants yet. All the mergers are triggered by three-body encounters. Binaries that disappear from Fig.~\ref{fig:fig8} and do not merge  split because of three-body encounters with single stars or because of SN explosions. 

At $t>4$ Myr, when the core  expands and the central density of the system decreases, the situation becomes simpler. Some of the PCPs, which at $t\sim{}3-7$ Myr have become BHs (Fig.~\ref{fig:fig5}), were able to retain a binary companion. These binaries are stable in most cases, and survive for the entire simulation. In contrast, those PCPs that did not retain a companion after the early stages do not form stable binaries anymore, because the core density is lower. 

Most importantly, all stable binaries containing a PCP are composed of two compact objects. At $Z=0.01$ Z$_\odot$ four PCPs out of ten form stable BH-BH binaries, with periods ranging from $\sim{}0.07$ yr to $\sim{}5$ yr. Two PCPs form two unstable BH-BH binaries, which break  by three-body encounters  before the end of the simulation. Finally, one PCP forms a stable binary with a NS, with a period of $\sim{}6$ days. 

At $Z=0.1$ Z$_\odot$ the PCPs form two stable BH-BH binaries, with period 0.6 and 3.7 yr, respectively, plus one extremely unstable BH-BH binary. Similarly, at $Z=$ Z$_\odot$ the PCPs form two stable BH-BH binaries, with period 9.7 and 19 yr, respectively, plus one extremely unstable BH-BH binary.   At higher metallicities, the formation of stable binaries is less frequent, probably because the mass of the PCP is smaller if compared to the other objects in the YDSC. In fact, only binaries whose members are significantly more massive than the other stars are stable against dynamical exchanges with intruders (see e.g. \citealt{hills1980}). Moreover, at low metallicity many SNe are failed, and the BHs form without natal kick, while natal kicks tend to unbind binaries at high metallicity (see  Section.~\ref{sec:structure}).  %binaries are not needed to keep the YDSC core stable against collapse at high metallicity, because stellar winds  are very effective, and three-body encounters do not need to transfer as much kinetic energy as in the metal-free case \citep{mapellibressan2013,trani2014}.}

 Table~\ref{tab:table2} summarizes the properties of the stable binaries at the end of the simulations. 

%%%%%%%%%%%%%%%%%%%%%%%%%%%%%%%%%%% FIGURE 8 %%%%%%%%%%%%%%%%%%%%%%%%%%%%%%%%%%
%%%%%%%%%%%%%%%%%%%%%%%%%%%%%%%%%%%%%%%%%%%%%%%%%%%%%%%%%%%%%%%%%%%%%%%%%%%%%%%

%%%%%%%%%%%%%%%%%%%%%%%%%%%%%%%%%%% FIGURE 9 %%%%%%%%%%%%%%%%%%%%%%%%%%%%%%%%%%
\begin{figure}
\center{{
\epsfig{figure=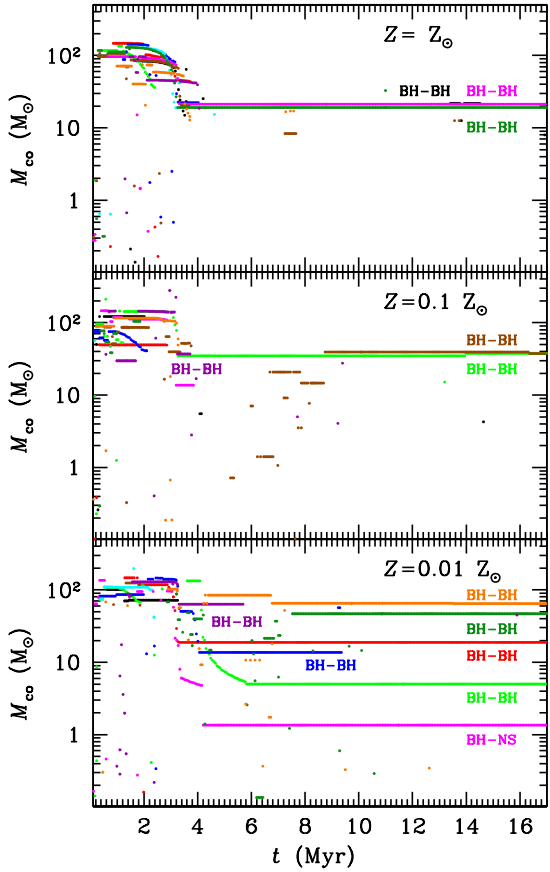,width=7.5cm} %PCP_Mco.eps
}}
\caption{\label{fig:fig9}
Mass of the companion of the PCP as a function of time, when the PCP is in a binary system. From top to bottom: $Z=$ Z$_\odot$, $Z=0.1$ Z$_\odot$ and $Z=0.01$ Z$_\odot$. Each line is a single PCP. The labels 'BH-BH' and 'BH-NS' mark PCPs that are members of 'black hole-black hole' and 'black hole-neutron star' binaries, respectively. Colours are the same as in Fig.~\ref{fig:fig8}. 
}
\end{figure}
%%%%%%%%%%%%%%%%%%%%%%%%%%%%%%%%%%%%%%%%%%%%%%%%%%%%%%%%%%%%%%%%%%%%%%%%%%%%%%%

%%%%%%%%%%%%%%%%%%%%%%%%%%%%%%%%%%% FIGURE 10 %%%%%%%%%%%%%%%%%%%%%%%%%%%%%%%%%%
\begin{figure}
\center{{
\epsfig{figure=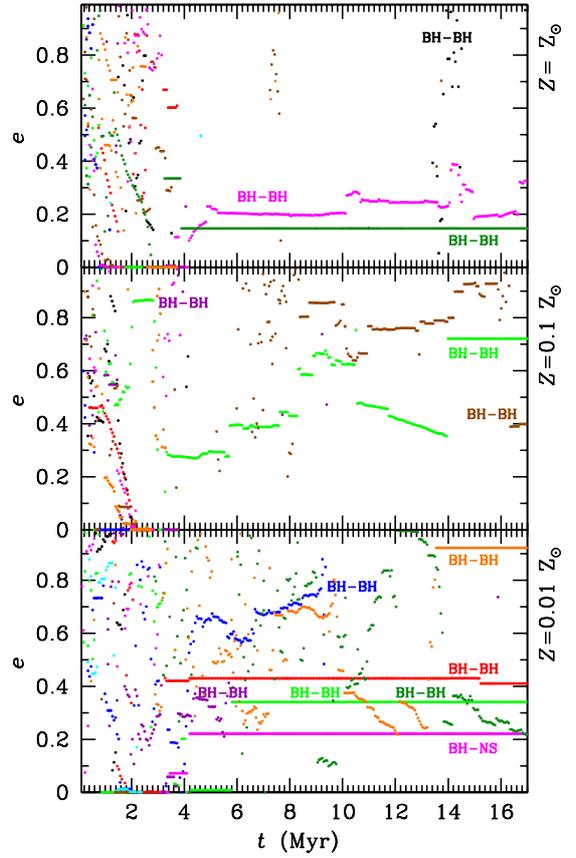,width=8cm} %eccPCP.eps
}}
\caption{\label{fig:fig10}
Eccentricity of the binary systems whose member is a PCP as a function of time. From top to bottom: $Z=$ Z$_\odot$, $Z=0.1$ Z$_\odot$ and $Z=0.01$ Z$_\odot$. Each line is a single PCP. The labels 'BH-BH' and 'BH-NS' mark PCPs that are members of 'black hole-black hole' and 'black hole-neutron star' binaries, respectively. Colours associated with the single runs are the same as in Fig.~\ref{fig:fig8}.
}
\end{figure}
%%%%%%%%%%%%%%%%%%%%%%%%%%%%%%%%%%%%%%%%%%%%%%%%%%%%%%%%%%%%%%%%%%%%%%%%%%%%%%%

Fig.~\ref{fig:fig9} shows the mass of the companion of the PCP ($M_{\rm co}$) as a function of time. In the first $\sim{}4$ Myr the companions are massive MS stars (with some exception of PCPs bound to low-mass MS stars) and  lose mass by stellar winds. At later stages, the companions are mostly dark remnants.  At $Z=$ Z$_\odot$, the two stable binaries are composed of two BHs with masses (20,21) M$_\odot$ and (5,19) M$_\odot$  in runs n6 and n8, respectively. At $Z=0.1$ Z$_\odot$, the two stable binaries are composed of two BHs with masses (19,38) M$_\odot$ and (254,38) M$_\odot$  in runs n3 and n7, respectively. At $Z=0.01$ Z$_\odot$, we find a variety of binary masses, ranging from (16,1.36) M$_\odot$ (in the case of the BH-NS binary, run n6) to (212,47) M$_\odot$ (in run n8). 

We note that three BH-BH binaries underwent an exchange in the late stages of the YDSC evolution. These are: (i) the BH-BH binary in run n10 at $Z=0.01$ Z$_\odot$ (orange line), in which  $M_{\rm co}$ changes from 65 to 64 M$_\odot{}$ at $t=13.6$ Myr; (ii) the BH-BH binary in run n3 at $Z=0.1$ Z$_\odot$ (light green line),  in which  $M_{\rm co}$ changes from 34 to 38 M$_\odot{}$ at $t=14$ Myr, and (iii) the BH-BH binary in run n7 at $Z=0.1$ Z$_\odot$ (brown line), in which  $M_{\rm co}$ changes from 39 to 38 M$_\odot{}$ at $t=16.3$ Myr. In all cases,  both the former and the new companions are BHs. %All the exchanges lead to an increase of eccentricity.

Fig.~\ref{fig:fig10} shows the eccentricity of the PCP binaries as a function of time. The eccentricity changes very fast  in many systems, even in  some of the most stable PCP binaries. In case of double compact-object binaries, changes of eccentricity are entirely due to exchanges and other dynamical interactions.  Highly eccentric systems are associated with recent exchanges (as in the cases of run n3 at $Z=0.1$ Z$_\odot$ and run n10 at $Z=0.01$ Z$_\odot$, see the light green line in the central panel and the orange line in the bottom panel of Figs.~\ref{fig:fig9} and \ref{fig:fig10}, respectively), and with strong perturbations induced by an intruder (as in the case of run n7 at  $Z=0.1$ Z$_\odot$, see the brown line in the central panel of Figs.~\ref{fig:fig9} and \ref{fig:fig10}; the intruder exchanges with the companion of the PCP at the end of the simulation). 

Finally, only the BH-NS binary (run n6 at $Z=0.01$ Z$_\odot$) and one of the eight BH-BH binaries (run n8 at $Z=$Z$_\odot$) were ejected from the parent YDSC during the simulations, while the other seven BH-BH binaries are still members of the YDSC.

%Table~\ref{tab:table2} summarizes the properties of the stable binaries at the end of the simulations.

 %%%%%%%%%%%%%%%%%%%%%%%%%%%%%%%%%% TABLE 2%%%%%%%%%%%%%%%%%%%%%%%%%%%%%%%%%%%%%
\begin{table*}
\begin{center}
\caption{\label{tab:table2}
Properties of the stable PCP binaries at the end of the simulations.}
 \leavevmode
\begin{tabular}[!h]{cccccccccc}
\hline
Run &  $Z/$Z$_\odot$   & $M_{\rm PCP}/{\rm M}_\odot$ & $M_{\rm co}/{\rm M}_\odot$ & $m_{\rm c}/{\rm M}_\odot$ & $P_{\rm orb}$/yr & $e$ & Type & Ejected? & $t_{\rm GW}/10^{10}{\rm yr}$ \\
\hline
n2   & 0.01   & 32  & 19   & 21 & 1.63   & 0.41 & BH-BH & no  & $3.9\times{}10^5$\\
n3   & 0.01   & 22  & 5    & 9 & 0.0685 & 0.34 & BH-BH & no  & $4.6\times{}10^{2}$\\ 
n6   & 0.01   & 16  & 1.36 & 4 & 0.0160 & 0.22 & BH-NS & yes & $5.4\times{}10^{1}$ \\
n8   & 0.01   & 212 & 47   & 82 & 4.50   & 0.35 & BH-BH & no  & $9.7\times{}10^5$\\
n10  & 0.01   & 135 & 64   & 80 & 1.37   & 0.92 & BH-BH & no  & $6.6\times{}10^{1}$\\
n3   & 0.1    & 19  & 38   & 23 & 3.67   & 0.72 & BH-BH & no  & $4.3\times{}10^{5}$\\
n7   & 0.1    & 254 & 38   & 79 & 0.60   & 0.39 & BH-BH & no  & $3.2\times{}10^{3}$\\
n6   & 1.0    & 20  & 21   & 18 & 9.67   & 0.31 & BH-BH & no  & $7.9\times{}10^7$\\
n8   & 1.0    & 5   & 19   & 8  & 19.3   & 0.15 & BH-BH & yes & $2.6\times{}10^9$\\
\noalign{\vspace{0.1cm}}
\hline
\end{tabular}
\begin{flushleft}
  \footnotesize{Column 1: run number; column 2: metallicity ($Z$); column 3: mass of the PCP ($M_{\rm PCP}$); column 4: mass of the companion ($M_{\rm co}$); column 5: chirp mass ($m_{\rm c}$); column 6: orbital period at the end of the simulation ($P_{\rm orb}$); column 7: eccentricity at the end of the simulation ($e$); column 8: type of binary members; column 9: status of the binary (`yes' means ejected from the YDSC, `no' means retained); column 10: coalescence timescale ($t_{\rm GW}$) due to GW emission \citep{peters1964}.}
\end{flushleft}
\end{center}
\end{table*}
%%%%%%%%%%%%%%%%%%%%%%%%%%%%%%%%%%%%%%%%%%%%%%%%%%%%%%%%%%%%%%%%%%%%%%%%%%%%%%%%

%%%%%%%%%%%%%%%%%%%%%%%%%%%%%%%%%%% FIGURE 11 %%%%%%%%%%%%%%%%%%%%%%%%%%%%%%%%%%
\begin{figure}
\center{{
\epsfig{figure=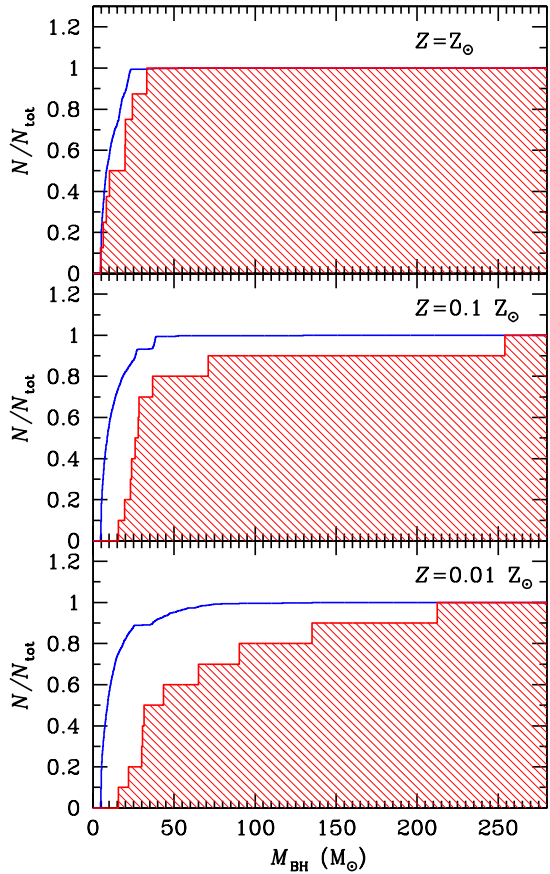,width=7.5cm} %was wrt_otherBH.eps
}}
\caption{\label{fig:fig11}
Cumulative mass distribution of the BHs born from the PCPs (red shaded histogram) versus the one of all BHs in the simulations (blue empty histogram). From top to bottom: $Z=$ Z$_\odot$, $Z=0.1$ Z$_\odot$ and $Z=0.01$ Z$_\odot$. 
}
\end{figure}
%%%%%%%%%%%%%%%%%%%%%%%%%%%%%%%%%%%%%%%%%%%%%%%%%%%%%%%%%%%%%%%%%%%%%%%%%%%%%%%

%%%%%%%%%%%%%%%%%%%%%%%%%%%%%%%%%%% FIGURE 12 %%%%%%%%%%%%%%%%%%%%%%%%%%%%%%%%%%
\begin{figure}
\center{{
\epsfig{figure=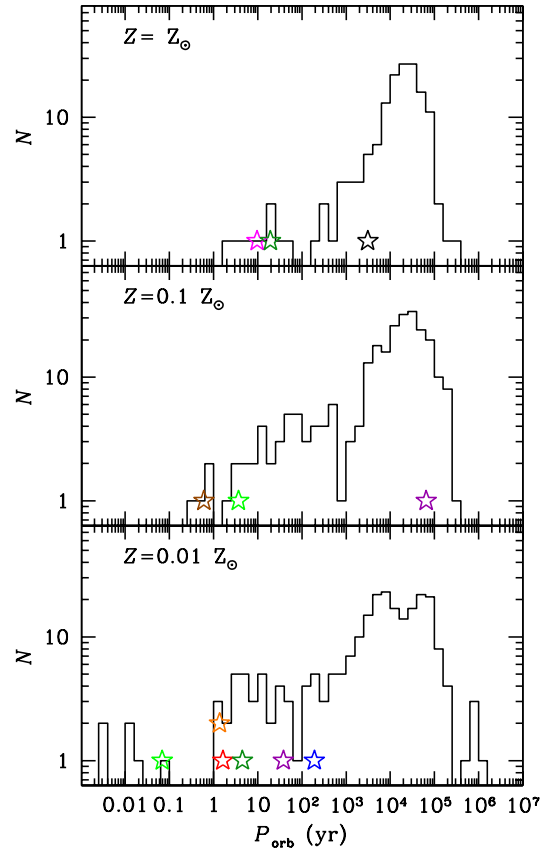,width=7.5cm} %was wrt_otherBH_period.eps fig11.eps
}}
\caption{\label{fig:fig12}
Distribution of minimum orbital periods of all BH-BH binaries in the simulations. Coloured stars indicate the minimum period of a BH-BH binary hosting a PCP (colours are the same as in Fig.~\ref{fig:fig8}). From top to bottom: $Z=$ Z$_\odot$, $Z=0.1$ Z$_\odot$ and $Z=0.01$ Z$_\odot$. 
}
\end{figure}
%%%%%%%%%%%%%%%%%%%%%%%%%%%%%%%%%%%%%%%%%%%%%%%%%%%%%%%%%%%%%%%%%%%%%%%%%%%%%%%

%%%%%%%%%%%%%%%%%%%%%%%%%%%%%%%%%%% FIGURE 13 %%%%%%%%%%%%%%%%%%%%%%%%%%%%%%%%%%
\begin{figure}
\center{{
\epsfig{figure=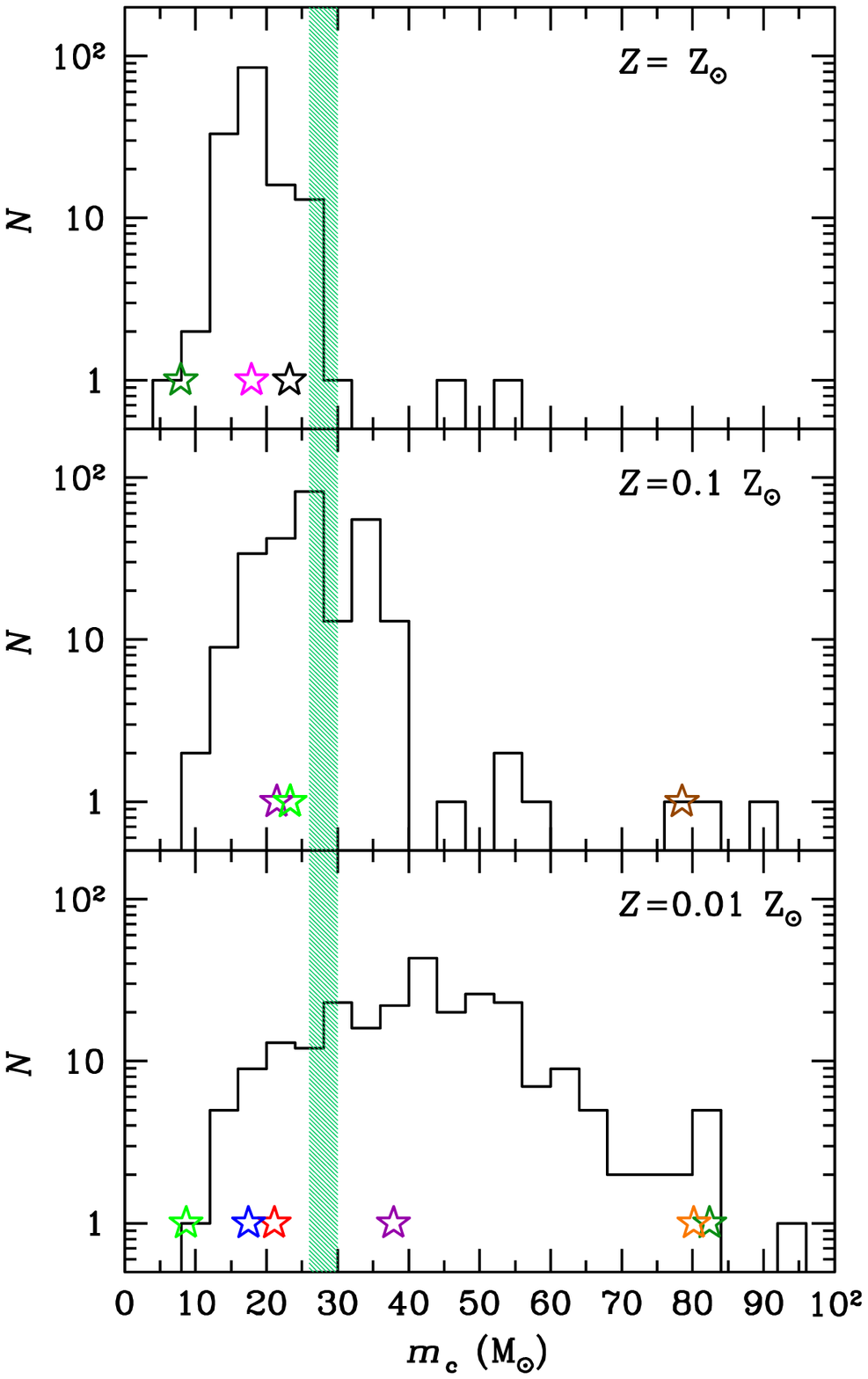,width=7.5cm} %was mchirpok.eps
}}
\caption{\label{fig:fig13}
Distribution of chirp masses ($m_{\rm c}$) of all BH-BH binaries in the simulations. Most of these binaries are unstable. Coloured stars indicate the final chirp mass of a BH-BH binary hosting a PCP (colours are the same as in Fig.~\ref{fig:fig12}). The shaded green area is the measured chirp mass of GW150914 \citep{LIGO2016}. From top to bottom: $Z=$ Z$_\odot$, $Z=0.1$ Z$_\odot$ and $Z=0.01$ Z$_\odot$. 
}
\end{figure}
%%%%%%%%%%%%%%%%%%%%%%%%%%%%%%%%%%%%%%%%%%%%%%%%%%%%%%%%%%%%%%%%%%%%%%%%%%%%%%%
\subsection{Comparison with other BHs in the YDSC}
As we discussed in the previous sections, most PCPs evolve into BHs. Some of them form double-compact object binaries. Is the population of BHs born from PCPs different from the rest of the BH population in the simulated YDSCs?

Fig.~\ref{fig:fig11} shows the cumulative distribution of BH masses in all simulated YDSCs compared to the cumulative distribution of masses of BHs born from PCPs. The most massive BH born from a PCP is always the most massive BH for a given  metallicity. The mass distribution of BHs born from PCPs is significantly skewed toward higher masses than the distribution of all BHs. The Kolmogorov-Smirnov (KS) test probability that the overall BH distribution and the distribution of BHs born from PCPs are drawn from the same population is $3\times{}10^{-5}$, $3\times{}10^{-4}$ and $10^{-3}$ for $Z=0.01,0.1$ and 1 Z$_\odot$, respectively. Thus, the three distributions are significantly different. %However, by construction, there are 

Many BH-BH binaries form in the simulated YDSCs dynamically:  246, 257 and 153 BH-BH binaries at $Z=0.01,$ 0.1 and 1 Z$_\odot$ (the efficiency is lower at solar metallicity because of the smaller average BH masses). Fig.~\ref{fig:fig12} compares the distribution of the minimum orbital period of all simulated BH-BH binaries with the orbital periods of BH-BH binaries that host a PCP.  Most BH-BH binaries are extremely soft, i.e. their orbital period is extremely long for the entire simulation. As already shown by \cite{ziosi2014}, such soft BH-BH binaries have short lifetimes, and continuously break and change members by dynamical exchanges.  The shortest period of a BH-BH binary at low metallicity is $\sim{}100$ times shorter than the shortest period at high metallicity. %This is an effect of the stronger dynamical activity of binary systems at low metallicity, where stellar winds do not remove mass from the core of the YDSC. 
The stable BH-BH binaries that contain a PCP are among the BH-BH binaries with the shortest period, at all metallicities.
%If we look at the properties of all BHs that form a binary with another BH....

%**AGGIUNGERE FIGURE PER ECCENTRICIT\`A
 Fig.~\ref{fig:fig13} shows the distribution of the chirp mass of the BH-BH binaries, defined as $m_{\rm c}=(m_1\,{}m_2)^{3/5}(m_1+m_2)^{-1/5}$, where $m_1$ and $m_2$ are the masses of the two components of the binary. The chirp mass is particularly important for GWs, since it determines how fast the binary sweeps, or chirps, through a frequency band (the amplitude and the frequency of GWs scale as $m_{\rm c}^{5/3}$ and $m_{\rm c}^{-5/8}$, respectively).  %As in Fig.~\ref{fig:fig12}, stars mark the BH-BH binaries that contain PCPs. 
The distribution of chirp masses strongly depends on the metallicity, as a consequence of the assumed recipes for BH formation. We note that several PCP binaries have relatively small chirp masses, and do not differ from the rest of the sample.  %, because the chirp mass depends not only on the PCP mass but also on the mass of the companion.}

%%%%%%%%%%%%%%%%%%%%%%%%%%%%%%%%%%% FIGURE 14 %%%%%%%%%%%%%%%%%%%%%%%%%%%%%%%%%%
\begin{figure}
\center{{
\epsfig{figure=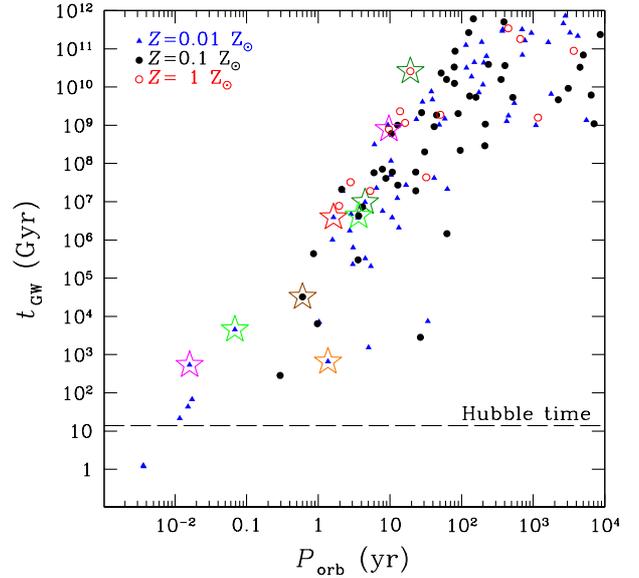,width=8cm} %was GWtime.eps 
}}
\caption{\label{fig:fig14}
Coalescence timescale of BH-BH and BH-NS binaries in the simulated  YDSCs as a function of their orbital period. Most simulated BH binaries are not shown in this plot because they have $t_{\rm GW}>10^{12}$ Gyr or $P_{\rm orb}>10^4$ yr. Blue filled triangles: BH binaries at $Z=0.01$ Z$_\odot$; black filled circles: BH binaries at $Z=0.1$ Z$_\odot$; red open circles: BH binaries at $Z=$ Z$_\odot$. The nine stars mark the position of the eight stable BH-BH binaries and of the NS-BH binary that contain a PCP (colours are the same as in Fig.~\ref{fig:fig2}).  The horizontal black dashed line corresponds to the Hubble time.
}
\end{figure}
%%%%%%%%%%%%%%%%%%%%%%%%%%%%%%%%%%%%%%%%%%%%%%%%%%%%%%%%%%%%%%%%%%%%%%%%%%%%%%%

\section{Discussion}
\subsection{Predictions for gravitational wave events}
The first direct detection of GWs was  recently reported by \cite{abbott2016}. The signal has been interpreted as emitted by a merging BH-BH binary with masses ($36^{+5}_{-4}$, $29^{+4}_{-4}$) M$_\odot$  \citep{LIGO2016b}. Previous studies already predicted the existence of such massive stellar  BHs (e.g. \citealt{mapelli2009,belczynski2010,spera2015}). However, %population synthesis simulations suggest that 
 it is difficult that an isolated binary system composed of two massive stars evolves into a BH-BH binary with such massive BHs (e.g. \citealt{linden2010}), because  mass transfer and common envelope  often lead to the merger of the two stars or to the loss of most of the initial mass of the system.  

In YDSCs, a massive BH-BH binary might form from dynamical exchanges \citep{ziosi2014}. In this case, the two massive BHs did not originate in the same primordial binary, but they entered the same binary as a consequence of dynamical exchanges with other stars.  

\cite{ziosi2014} find that most BH-BH mergers in YDSCs   are expected to occur between low-mass BHs ($5-15$ M$_\odot$),  despite  the efficiency of dynamical exchanges (see also  \citealt{downing2010,downing2011,oleary2006,sadowski2008,clausen2013,morscher2015,rodriguez2015,amaro2015,oleary2016,chatterjee2016} for other predictions). The YDSCs simulated by \cite{ziosi2014}  have initial mass $\sim{}3500$ M$_\odot$ (i.e. about 20 times less massive than the YDSCs we simulated). Do the results of \cite{ziosi2014} still hold in more massive and denser YDSCs? 

%Advanced LIGO 

In the previous Section, we reported that several PCPs form stable BH-BH or even BH-NS binaries in the simulated YDSCs. Two of such binaries have masses  similar to  the merging binary system that was recently detected by Advanced LIGO. These are system n2 at $Z=0.01$ Z$_\odot{}$ and system n3 at $Z=0.1$ Z$_\odot{}$, which have mass (32,19) and (38,19) M$_\odot$, respectively.

We now estimate the coalescence timescale for such PCP binaries as well as for the other BH-BH binaries in the simulations. We define the coalescence timescale as the time for the semi-major axis of a binary to sink to zero, due to the emission of GWs \citep{peters1964}
\begin{equation}
t_{\rm GW}=\frac{5}{256}\,{}\frac{c^5\,{}a^4\,{}(1-e^2)^{7/2}}{G^3\,{}m_1\,{}m_2\,{}(m_1+m_2)},
\end{equation}
where $c$ is the speed of light, $G$ is the gravitational constant, $a$ and $e$ are the semi-major axis and the eccentricity of the binary system composed of two objects with mass $m_1$ and $m_2$.

Fig.~\ref{fig:fig14} shows $t_{\rm GW}$ for the simulated BH binary systems that are closest to coalescence, including the nine stable PCP binaries in Table~\ref{tab:table2}. Apart from the BH-NS binary already discussed in this paper, all the other systems are BH-BH binaries, indicating that the formation of BH-NS binaries is extremely rare if primordial binaries are not included.  For the same reason, we decided not to include the statistics of NS-NS binaries. \cite{ziosi2014} recently showed that nearly all NS-NS binaries come from primordial binaries, even in YDSCs. 

All but one simulated double-compact object systems are not expected to coalesce within a Hubble time. The only BH-BH binary that is expected to coalesce within the Hubble time (with $t_{\rm GW}\sim{}1.2$ Gyr) is at low metallicity ($Z=0.01$ Z$_\odot$) and has a mass of (17,\,{}16) M$_\odot$, a minimum orbital period $P_{\rm orb}\sim{}1.3$ days and an eccentricity $e\sim{}0.061$. There is a clear trend with metallicity: the BH systems with the shorter $t_{\rm GW}$ are in metal-poor YDSCs. The reason is that metal-poor YDSCs host more massive BHs (which are more efficient in acquiring companions through dynamical interactions, \citealt{ziosi2014}). % and undergo more dynamical interactions (to keep the core stable against collapse in absence of stellar winds, e.g. \citealt{mapellibressan2013}).

All nine stable PCP binaries are among the systems with the shortest $t_{\rm GW}$, but they are not expected to merge within a Hubble time. It must be said, however, that the orbital properties of these binaries are expected to change if they remain in the YDSC after the end of the simulation. Three-body encounters (especially exchanges) can further reduce the semi-major axis and increase the eccentricity, leading to a decrease of $t_{\rm GW}$, too. For example, from Fig.~\ref{fig:fig8} it is apparent that the period of several BH-BH binaries is still decreasing because of three-body encounters and dynamical exchanges. In this paper, we chose to stop the simulations at $t=17$ Myr because we did not add any prescription for the tidal field of the host galaxy and we cannot account for tidal evaporation of the YDSC. In a forthcoming study, we will add recipes for the tidal field of the host galaxy.  

Our results seem to confirm the main finding of \cite{ziosi2014}: while most BH-BH binaries (and especially the most massive BH-BH binaries) in YDSCs form from dynamical exchanges, it is difficult that dynamically formed systems merge within a Hubble time, even if they host the product of runaway collisions.  Our result holds if YDSCs live for $<10^8$ Myr. Thus, the time when YDSCs are disrupted in the tidal field of the host galaxy is important.  However, we must stress that the simulated sample (30 YDSCs) is statistically small. Using the same formalism as in equation~3 of \cite{ziosi2014}, the predicted merger rate of the simulated BH-BH systems is
\begin{eqnarray}\label{eq:rate}
R_{\rm BH-BH}\sim{}10^{-3}\,{}{\rm Mpc}^{-3}\,{}{\rm Myr}^{-1}\,{}\,{}\left(\frac{n_{\rm merg}}{1}\right)\,{}\left(\frac{2\times{}10^6{\rm M}_\odot}{M_{\ast}}\right)\nonumber\\\,{}\left(\frac{1.2\,{}{\rm Gyr}}{t_{\rm GW}}\right)\,{}\left(\frac{t_{\rm life}}{10^8{\rm yr}}\right)\,{}\left(\frac{\rho{}_{\rm SFR}}{0.015\,{}{\rm M}_\odot{}{\rm yr}^{-1}\,{}{\rm Mpc}^{-3}}\right),
\end{eqnarray}
where $n_{\rm merg}$ is the number of BH-BH systems that merge in a Hubble time in our simulations, $t_{\rm GW}$ is the maximum coalescence time of such systems,  $M_\ast$ is the total simulated stellar mass (given by the number of simulated YDSCs multiplied by their mass),  $t_{\rm life}$ is the lifetime of an YDSC, and $\rho{}_{\rm SFR}$ is the cosmic star formation rate density at low redshift \citep{hopkins2006}. Since the instrumental range of Advanced LIGO and Virgo for BH-BH binaries with chirp mass $m_{\rm c}\sim{}15$ M$_\odot{}$ is $\sim{}1$ Gpc, equation~\ref{eq:rate} leads to a detection rate $\approx{}4$ yr$^{-1}$.  %{\bf To derive the estimate in equation~\ref{eq:rate} we assumed that the merger rate scales with the cosmic star formation rate and we neglected metallicity differences in the simulated YDSCs.

 The minimum merger rate derived from the LIGO detection is $\sim{}2\times{}10^{-3}$ Mpc$^{-3}$ Myr$^{-1}$ \citep{LIGO2016b}, i.e. similar to (and slightly higher than) what we found.  However, the merger rate we obtain in equation~\ref{eq:rate} should be taken as a strong lower limit, because we neglect the contribution of primordial binaries. To get more accurate estimates, we need a larger simulation sample, and we must include primordial binaries. 

 %For this reason and because of the low statistics, this result can be considered only an order-of-magnitude estimate. 

To conclude, we point out that the most massive BH-BH binaries and the best merger candidates are at metallicity $Z=0.01$ Z$_\odot$. This metallicity is typical of old Milky Way globular clusters. If the properties of globular clusters in the early stages of their evolution were similar to the properties of YDSCs in the nearby Universe, this means that the most massive BH-BH binaries formed in globular clusters. These massive BH-BH binaries born in globular clusters can give an important contribution to detectable GW signal. %Our simulations show that the majority of these BH-BH binaries remain in their birthplace.

%le nostre assunzioni sono ottimistiche, mma....

%dire che dopo la detection di LIGO Virgo il problema non \`e tanto spiegare come fanno a esistere  i BH massivi (mapelli,spera.....) ma piuttosto come fanno 2 BH massivi a trovarsi in binaria assieme. il common envelope etcetc...mentre la dinamica etcetc...

%calcoliamo il rate

%\caption{\label{fig:fig2}
%Radial distance of the satellite from the centre of the primary galaxy for all simulations, as a function of time. Time $t=0$ corresponds to the first periapsis passage. Long-short dashed black line: run~A; solid red line: run~B; dotted ochre line: run~C; dashed green line: run~D; long-dash-dotted blue line: run~E; short-dash-dotted magenta line: run~F.
%}
%\end{figure}
%%%%%%%%%%%%%%%%%%%%%%%%%%%%%%%%%%%%%%%%%%%%%%%%%%%%%%%%%%%%%%%%%%%%%%%%%%%%%%%
%\section{Discussion}
\subsection{Comparison with other scenarios of IMBH formation}
  The runaway merger scenario is one of the principal mechanisms that were proposed to form IMBHs. Other popular scenarios of IMBH formation are the direct collapse of very massive ($>250$ M$_\odot$) population III stars \citep{madau2001,schneider2002,heger2003}, and the repeated mergers of stellar-mass BHs with stars and other BHs, triggered by dynamical interactions in star clusters \citep{miller2002,giersz2015}.

Both the latter model and the runaway merger scenario rely on the importance of dynamical interactions to trigger the formation of IMBHs. The main difference between these two models is the formation timescale: the runaway merger occurs very fast (mostly before the collapse of the PCP into a BH), while the scenario of repeated mergers involves objects that have already collapsed to BHs. However, the distinction between these two models is rather weak, since even in the runaway merger scenario an IMBH can  keep growing through mergers with other BHs after its formation. For example, \cite{giersz2015} note that, in their simulations, there is a transition between a first epoch of fast formation of IMBHs by collisions (at the beginning of cluster evolution) and a second epoch (at $t\gtrsim{}3$ Gyr) of slower build-up of IMBHs by dynamical encounters involving binary systems. Our results are reminiscent of the epoch of fast IMBH formation described in \cite{giersz2015}, while the fact that our simulated IMBHs tend to form binaries with other BHs suggests that they might undergo more binary interactions at later times, as studied by \cite{giersz2015}. 

In contrast, IMBHs formed by the collapse of metal-free stars at high redshift are not expected to be in star clusters nowadays. As shown by previous studies \citep{mapelli2006,mapelli2007,kuranov2007}, it is very unlikely for these population III IMBHs to acquire a companion and to be observed as sources of X-ray  emission or GWs.

\subsection{Caveats and future work}
 Unraveling the runaway merger in YDSCs is an arduous task, because it involves a plethora of astrophysical processes (e.g. dynamical friction, mass segregation, dynamics of three-body encounters, stellar and binary evolution, hydrodynamics of collisions, BH formation, structural star cluster evolution), that occur on considerably different scales (from the stellar radius up to several parsecs). Hence, many approximations and assumptions were needed, in order to make our simulations feasible. In this Section, we discuss the main issues of these approximations.

We assumed that the entire mass of two colliding objects (either stars of BHs) ends up into the collision product. Hydrodynamical simulations of the collision of massive stars indicate that mass losses can be of the order of $\sim{}25$\% of the entire mass of the merger product. Moreover, if the two colliding objects are a star and a BH, less than 50\% of the incoming star mass might be directly accreted onto the BH (see e.g. \citealt{ramirez2009,guillochon2013}). This implies that the maximum masses of PCPs shown in our Fig.~\ref{fig:fig5} are overestimated. This issue is particularly sensitive for collisions of stars with BHs, because the collision product does not undergo further mass loss by stellar winds. In their Monte Carlo simulations, \cite{giersz2015} compare simulations where 100\% and 25\% of the mass of the colliders end up in the collision product, respectively. When only 25\% of the mass is accreted onto the collision product, the final mass is significantly lower, but IMBH formation is still possible (see fig.~7 of \citealt{giersz2015}).

To evolve the most massive stars, we extend the fitting formulate of \cite{hurley2000} to masses $>100$ M$_\odot$. While we use PARSEC stellar evolution tracks to check that the radii do not become unphysical \citep{bressan2012,tang2014,chen2015}, this assumption is not self-consistent. In future, we plan to use up-to-date PARSEC stellar evolution tracks for all stars (Spera et al., in preparation). Moreover, we used an old prescription for stellar rejuvenation \citep{portegieszwart1999}, which should be updated accounting for chemical mixing.

No primordial binaries are included in our simulations. The reason is that integrating binaries is the bottleneck of direct N-body simulations, but we know that a high binary fraction is observed in young star clusters (e.g. \citealt{li2013}). If there are no primordial binaries, binaries can form through dynamical encounters. This mechanism is efficient in our simulations (see Fig.~\ref{fig:fig1}), thanks to the high central density, but the absence of primordial binaries has an impact on our results. \cite{mapelli2013} and \cite{mapelli2014} study the effect of different primordial binary fractions on the demographics of BH binaries (defined as binaries hosting at least one BH) in YDSCs with  mass $M_{\rm TOT}\sim{}3500$ M$_\odot$, i.e. 20 times smaller than the YDSCs we simulate here. They assume that 0, 18 and 33 \% of stars are members of primordial binaries in their simulations, and find that the number of BH binaries born from a dynamical exchange does not depend on the primordial binary fraction, whereas the number of BH binaries born from primordial binaries scales almost linearly with the primordial binary fraction. This implies  that our merger rate of BH-BH binaries is a strong lower limit. %The star clusters simulated by \cite{mapelli2013}, \cite{mapelli2014} and \cite{ziosi2014} are too small for a runaway merger to occur. Thus, we cannot use their results to infer information about the impact of primordial binaries on the PCP mass. 
Moreover, Monte Carlo simulations \citep{leigh2015,giersz2015}, where a high fraction of primordial binaries can be accounted for, indicate that binary interactions are important for the build-up of IMBHs, by enhancing stellar collisions. %However, \cite{giersz2015} do not explicitly quantify the dependence of the IMBH mass on the primordial binary fraction.

Finally, our initial conditions do not account for the effects of gas from the parent molecular cloud. %This is a severe limitation. 
\cite{leigh2014} suggest that gas damping (i.e. the contribution of gas to dynamical friction) can accelerate mass segregation and even lead to cluster contraction. This might enhance the stellar collisions leading to the runaway merger. On the other hand, a relatively fast gas expulsion due to stellar winds and SN explosions can lead to the expansion of the cluster \citep{marks2008,marks2012,leigh2013}, reducing the efficiency of stellar collisions. This effect is evident also in our simulations: we do not include the gas of the parent molecular cloud, but mass loss by SNe and stellar winds leads to a significant expansion of the cluster, associated with a decrease of the total binary binding energy (Fig.~\ref{fig:fig1}). In a forthcoming study, we will account for the influence of gas on the early stages of the YDSC evolution.

\section{Conclusions}
 We investigated the runaway collision scenario by means of direct N-body simulations of YDSCs ($N_\ast=10^5$ particles) with three different metallicities: 0.01, 0.1 and 1 Z$_\odot$. We ran 10 realizations of the same YDSC per each metallicity, adopting a King model \citep{king1966}, with virial radius $r_{\rm v}=1$ pc and dimensionless central potential $W_0=9$. Stellar masses are distributed according to a Kroupa (2001) initial mass function.

%nostri risultati mostrano che è estremamente facile formare IMBH e MSBH se si va a basse metallicity e in ambienti poco densi
%Our assumptions for stellar collisions are optimistic, since 
 We assume no mass loss during the collision of two stars. Thus, the mass of the merger product immediately after each collision should be considered as an upper limit.  Moreover, we do not include primordial binaries, which might enhance stellar collisions, and we do not account for the influence of the relic gas from the parent molecular cloud.  %Another caveat about our simulations is that we adopt binary evolution recipes that were derived for lower-mass systems: in a forthcoming paper we will check and update binary evolution recipes.

 On the other hand, we adopt realistic prescriptions for the stellar winds \citep{mapelli2013}. Thus, we model mass loss by stellar winds self-consistently, and we account for the impact of this mass loss on subsequent collisions of each merger product with other stars, whereas  previous studies (e.g. \citealt{glebbeek2009}) included realistic stellar winds only {\it a posteriori},  in  N-body simulations that were run without stellar evolution. %in the post-processing of dynamical simulations.

The maximum mass that a PCP can achieve in our simulations is close to $\sim{}500$ M$_\odot$, regardless of the metallicity. The maximum  mass of the PCP is sensitive to the number of collisions and to the time when the first collision occurs. In contrast, the final mass of the PCP (when it becomes a stellar remnant) does not depend on the number of collisions and on the time of the first collision. The mass of the PCP remnant is affected by the metallicity of the progenitor stars, because stellar winds depend on metallicity. 

BHs with mass up to $\sim{}250$ M$_\odot$  form from the direct collapse of a PCP, if the metallicity is sufficiently low ($Z\le{}0.1$ Z$_\odot$) and if the central density of the YDSC is sufficiently high (i.e. in the first $\sim{}5$ Myr of the YDSC evolution). Three BHs born from PCPs have mass $\gtrsim{}90$ M$_\odot$, and can be considered genuine IMBHs. Moreover, if the metallicity is low ($Z\le{}0.1$ Z$_\odot$), massive stellar black holes ($25\le{}m_{\rm BH}/{\rm M}_\odot\le{}90$) can form from the direct collapse of a massive star, even without runaway collisions.

We find that $\sim{}60$ \% of the simulated PCPs are not ejected from their parent YDSC for the entire simulation. Most PCPs acquire companions through dynamical interactions. In the first $\sim{}4$ Myr, PCP binaries are short-lived: they are continuously destroyed by collisions, SN explosions and dynamical interactions. Those PCP binaries that remain bound after this early stage survive for the entire simulation. At the end of the simulations ($t=17$ Myr), all stable PCP binaries are double-compact object binaries. We find nine stable PCP binaries: eight of them are BH-BH binaries, and one is a BH-NS binary. % All of them but one (nine systems) are BH-BH binaries. 

Five of the nine stable PCP binaries (including the BH-NS binary) form at $Z=0.01$ Z$_\odot$. Among the other four systems, two BH-BH binaries are at $Z=0.1$ Z$_\odot{}$ and two are at $Z=$ Z$_\odot$. We suggest that double-compact object formation is more efficient at low metallicity, because the remnant mass is higher (massive objects are favoured in acquiring companions dynamically, \citealt{hills1991}). % and because the weakness of stellar winds enhances dynamical encounters \citep{mapellibressan2013}.  

The periods of the stable double-compact object PCP binaries range from few days up to $\sim{}20$ years.  Their eccentricities range from 0.15 to 0.92.  High eccentricities are associated with recent exchanges  and with strong perturbations induced by intruders. The masses of the primary members range  from $\sim{}15$ M$_\odot$ to $\sim{}250$ M$_\odot$ (Table~\ref{tab:table2}). Two of these binaries have masses  similar to the merging BH-BH binary system recently detected by Advanced LIGO. These are system n2 at $Z=0.01$ Z$_\odot{}$ and system n3 at $Z=0.1$ Z$_\odot{}$, which  have mass (32,19) and (38,19) M$_\odot$, respectively. In comparison with other simulated BH-BH binaries (those that do not contain the PCPs), the PCP binaries  have relatively short periods.  % and high masses}. %are among the most massive ones and the closest ones.
%The BHs born from the collapse

%PERIODS AND MASSES

%COMPARISON WITH OTHER BH

The merger rate of simulated BH-BH binaries is  $\sim{}10^{-3}$ Mpc$^{-3}$ Myr$^{-1}$, corresponding to $\sim{}4$ expected detections per year by Advanced LIGO and Virgo. %The minimum merger rate estimated on the basis of the first LIGO detection is $\sim{}2\times{}10^{-3}$ Mpc$^{-3}$ Myr$^{-1}$ \citep{LIGO2016b}, i.e. slightly higher than what we found. 
This predicted rate should be taken as a strong lower limit, because we neglect the contribution of primordial binaries.  None of the stable PCP binaries is expected to merge in a Hubble time. Increasing the statistics of simulated YDSCs  and adding a conspicuous fraction of primordial binaries are essential steps to make more accurate predictions for the detection of GWs from PCP binary mergers. 

%on the possibility of detecting GWs 

%, six out of ten simulated PCPBHs acquire a BH companion, forming a BH-BH binary. Four of these six binaries are stable and survive for the entire simulation. Moreover, one PCPBH acquires a NS companion, forming a stable BH-NS binary.
\section*{Acknowledgments}
I thank the referee, Mirek Giersz, for his accurate reading of the manuscript and for the numerous comments that improved this work significantly. I also thank Alessandro Bressan, Alessandro Trani, Emanuele Ripamonti, Marica Branchesi, Mario Spera, Ugo Niccol\`o Di Carlo, Enrico Montanari and Andrea Moretti for useful discussions. 
 I made use of the    {\sc starlab} (ver. 4.4.4) public software environment and of the {\sc SAPPORO} library \citep{gaburov2009sapporo}. I thank the developers of   {\sc starlab}, and especially its primary authors (P. Hut, S. McMillan, J. Makino, and S. Portegies Zwart). I thank the authors of {\sc SAPPORO}, and in particular E. Gaburov, S. Harfst, and S. Portegies Zwart. I acknowledge the CINECA Award N. HP10C3ANJY (VMStars) for the availability of high performance computing resources and support. The simulations were run on the graphics processing unit (GPU) cluster EURORA at CINECA. I acknowledge financial support from the Italian Ministry of Education, University and Research (MIUR) through grant FIRB 2012 RBFR12PM1F, from INAF through grant PRIN-2014-14, and from the MERAC Foundation.

\bibliography{./bibliography}
%\bibitem[\protect\citeauthoryear{Aguerri et al.}{2009}]{Aguerri2009}Aguerri J. A. L., M\'endez-Abreu J., Corsini E. M. 2009, A\&{}A, 495, 491
%\end{thebibliography}
%\onecolumn
\appendix
\section{Other collision products}
\begin{figure}
\center{{
\epsfig{figure=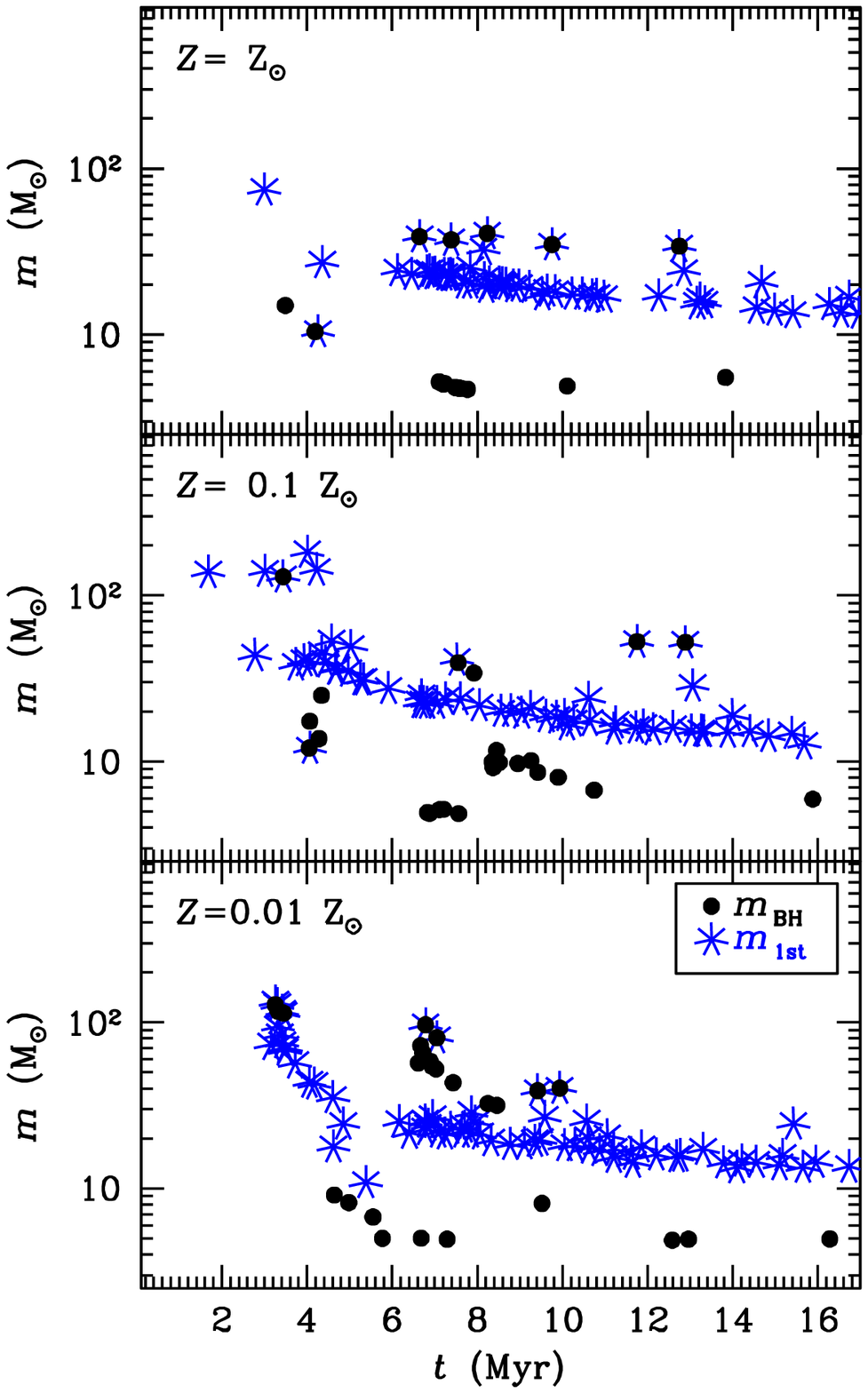,width=7.5cm} 
}}
\caption{\label{fig:figA1}
Mass of the OCP after the first collision ($m_{\rm 1st}$, blue asterisks), and mass of the BH born from the PCP ($m_{\rm BH}$, black filled circles) as a function of time for solar metallicity (top), $Z=0.1$ Z$_\odot$ (center) and $Z=0.01$ Z$_\odot$ (bottom). 
}
\end{figure}

 In this appendix, we summarize the properties of the other collision products that form after the PCP (hereafter, other collision products, OCPs). Fig.~\ref{fig:figA1} shows the masses of the OCPs and of the BHs born from the OCPs.
%%%%%%%%%%%%%%%%%%%%%%%%%%%%%%%%%%% FIGURE A1 %%%%%%%%%%%%%%%%%%%%%%%%%%%%%%%%%%
%%%%%%%%%%%%%%%%%%%%%%%%%%%%%%%%%%%%%%%%%%%%%%%%%%%%%%%%%%%%%%%%%%%%%%%%%%%%%%%

All the OCPs but two undergo a single collision during the entire simulation. No OCP undergoes more than two collisions. This is a striking difference with the PCPs, most of which undergo multiple collisions.  The maximum mass of the OCPs is always the same as the mass after the first collision ($m_{\rm 1st}$). %, which is the reason why we do not plot the maximum mass $m_{\rm max}$ in Fig.~\ref{fig:figA1} as we did in Fig.~\ref{fig:fig5}.

Fig.~\ref{fig:figA1} shows that the maximum mass of OCPs is always $<200$ M$_\odot$, and most OCPs have mass $<<100$ M$_\odot$. In general, the PCP is the most massive object in the star cluster. Moreover, the collision that leads to the formation of the first OCP is significantly delayed with respect to the formation of the PCP: all OCPs but three form at $t>3$ Myr. 

Despite this, some massive BHs form also from the collapse of the OCPs. Among the dark remnants born from the OCPs, three BHs at $Z=0.01$ Z$_\odot$ and one BH at $Z=0.1$ Z$_\odot$ have mass $>100$ M$_\odot$. 

%What
%is the mass distribution of collision products (other than PCP)? 
%always that the first collision leads to the formation of
%the most massive MS star (PCP) or maybe a more massive star can be formed in
%interactions which happen later? 

In our simulations, we find 69, 58, and 53 OCPs at $Z=0.01$, 0.1 and 1 Z$_\odot$, respectively. Only 26, 25 and 19 of these OCPs  become BHs at $Z=0.01$, 0.1 and 1 Z$_\odot$, respectively. These represent only the $\sim{}1.0-1.5$\% of all the BHs in the simulations (the total number of BHs in our simulations is 1689, 1746 and 1713 at $Z=0.01,$ 0.1 and 1 Z$_\odot$, respectively).  The mass of the other OCPs is too small to form a BH. Binary interactions are of primary importance for the formation of OCPs as they were for the formation of the PCP. This is consistent with previous results by e.g. \cite{gaburov2008a} and \cite{giersz2015}. 
%How frequent are collisions in which (other
%than PCP) massive MSs are formed and how many such collisions are present? 
%Is it always
%that binaries are involved in the formation of PCP? 
%Are binaries important for
%formation of other massive stars?
\end{document}